# "What It Wants Me To Say": Bridging the Abstraction Gap Between End-User Programmers and Code-Generating Large Language Models

Michael Xieyang Liu*
Microsoft Research,
Carnegie Mellon University
USA

Advait Sarkar*
Microsoft Research,
University of Cambridge,
University College London
UK

Carina Negreanu
Microsoft Research
UK

Benjamin Zorn
Microsoft Research
USA

Jack Williams
Microsoft Research
UK

Neil Toronto
Microsoft Research
UK

Andrew D. Gordon
Microsoft Research
UK

## ABSTRACT

Code-generating large language models map natural language to code. However, only a small portion of the infinite space of naturalistic utterances is effective at guiding code generation. For non-expert end-user programmers, learning this is the challenge of *abstraction matching*. We examine this challenge in the specific context of data analysis in spreadsheets, in a system that maps the user's natural language query to Python code using the Codex generator, executes the code, and shows the result. We propose *grounded* abstraction matching, which bridges the abstraction gap by translating the code back into a systematic and predictable naturalistic utterance. In a between-subjects, think-aloud study (n=24), we compare grounded abstraction matching to an ungrounded alternative based on previously established query framing principles. We find that the grounded approach improves end-users' understanding of the scope and capabilities of the code-generating model, and the kind of language needed to use it effectively.

---

*Equal contribution.

## 1 INTRODUCTION

Programming languages are an extremely powerful form of user interface. They also happen to be extremely difficult to learn, especially for non-expert end-user programmers who lack training in computing [48]. What if end-user programmers could instead use a natural language such as English? Natural language is already known to the user, and ostensibly requires little conscious investment of effort or learning. This prospect can be realized through large language models: deep neural networks using the transformer architecture [113], trained on large corpora, fine-tuned to generate code from natural language. For brevity, we use **LLM** to mean code-generating large language models. Despite impressive benchmark performance, LLMs are beset with issues in practical use. Lab and field studies have shown that the mapping between natural language and code is poorly understood, that generated code can contain subtle bugs, and that generated code can be difficult to verify [95, 112, 124].

In this paper, we consider the specific problem of *abstraction matching* [95]: when the user has a well-formed intent, how do they select an utterance from the near infinite space of naturalistic utterances that they believe the system will reliably map to a satisfactory solution? This involves "matching" the utterance to the right level of "abstraction", by specifying the utterance at a level of granularity and detail that matches the set of actions the system can take, and selecting suitable words and grammar.

The abstraction matching problem affects practically every natural language interface. Solutions (detailed in Section 2) include showing example commands, teaching users techniques such as breaking down their problem, operating with a restricted vocabulary and grammar [77], and incorporating other interface elements (e.g., graphical menus) to help users formulate their query. Each has drawbacks: examples are not necessarily reflective of user interests and do not help the user generalize to a wider range of utterances, tutorials take time, and restricted grammar reduces user flexibility.

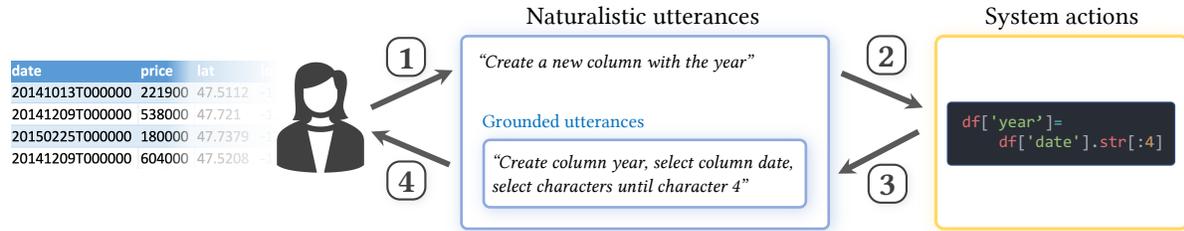

Figure 1: A summary of the user interaction loop in grounded abstraction matching. The user wishes to extract the year from a column of date strings. (1) The user chooses a naturalistic utterance to express their intent. (2) The utterance is mapped to a point in the space of system actions, in this case, a piece of Python code. (3) A grounded utterance is generated, which reflects the system action back to the user. (4) The user observes this utterance, interacts with it, and develops their mental model for future utterances.

We propose an alternative solution to the abstraction matching problem, which we call *grounded abstraction matching* (Section 3):

> An interface supports grounded abstraction matching if the user's naturalistic utterance is mapped to a system action, and then mapped back to a naturalistic utterance that is an editable example of how to consistently invoke the same action. This gives a *grounded* example of the level of abstraction at which the system expresses its solutions.

*The setting of our study.* Figure 1 is a sketch of the user interaction loop we propose. It depicts a concrete example of grounded abstraction matching in an interface for doing data analysis tasks in spreadsheets. The user selects an utterance (1) to express their intent from the large space of naturalistic utterances, such as *"Create a new column with the year"*. This is translated into the space of system actions, which in the example is the space of data analysis code (2). The code is then translated *back* into a smaller subspace of grounded utterances (3), which are in a consistent format, can reliably be interpreted by the system, and which we wish the user to learn. Finally (4), the user observes this grounded utterance (which influences their future use of the system) and can edit it to refine their query.

We investigate grounded abstraction matching in the context of end-user programmers solving data analysis tasks in spreadsheets. End-user programmers stand to benefit greatly from natural language programming, as they often do not have formal training in programming [48, 94]. On the other hand, their lack of programming expertise exacerbates the abstraction matching problem. Without knowledge of the underlying code generator and available APIs, it is much harder to formulate one's intent in terms that can be reliably translated to code. The combination of the high real-world value, as well as the acute interaction design challenge, makes this an ideal setting to study the problem. We make the following contributions in this paper:

- A description of the problem of abstraction matching situated in prior work, and a solution: grounded abstraction matching (Section 3). We show an example of how this solution can be instantiated in the context of end-user programmers solving data analysis tasks in spreadsheets (Section 4).
- We present a user study ($n = 24$) comparing grounded and ungrounded techniques for abstraction matching, showing that the grounded approach improves users' abilities to recover from system failures, and that users gain greater confidence and sense of control in using the system (Section 5-6).
- Our discussion throws new light on issues of end-user interaction with large language models. We find design tensions between prompt language and explanation language, and suggest that confusion of subtly different "dialects" in different natural language systems may be a future design challenge (Section 8).

## 2 BACKGROUND AND RELATED WORK

Natural language interaction faces many challenges. One of these challenges is abstraction matching: selecting an utterance that will translate into the desired system action. Various solutions have been proposed, but none appears to be targeted towards end-user programmers working with a large language model, or with the goal of helping users develop a mental model of the level of abstraction at which the LLM operates. Here, we review related problems and solutions identified in interaction design and machine learning research.

### 2.1 Natural language interfaces

Many usability issues of natural language interfaces can be seen as variations, or consequences, of Norman's "gulf of execution" [40]: the problem of getting the computer to do what you want it to do. The issues begin with forming an intent: a user intent may be beyond the capabilities of the system. Luger and Sellen [71] find a gulf between user expectations and the practical experience of conversational agents, suggesting that they should set realistic expectations to scaffold the learning process, and should reveal the system capabilities through interaction.

However, a well-formed intent is not enough; the user now faces the problem of targeting a set of system actions that might solve their problem (termed the *selection barrier* by Ko et al. [49]). For example, the user of a mobile phone voice assistant, while cooking, may form an intent not to let the egg overcook. This intent can be solved through the system action of setting a 2-minute timer. Design approaches to address the selection barrier include improving the visibility of the available actions, as in blocks programming languages [85], where actions are visualized and reified in a virtual toolbox. Similarly, "menu-driven" interfaces for natural language allow the user to construct naturalistic queries by pointing [108].

Even when a tool is selected, there are *use* barriers and *coordination* barriers (in Ko et al.'s terms), because the user needs to figure out how to operate the tool. In a programming language, this amounts to using syntax, APIs, data structures, etc. in the correct way [38, 39, 67–69], and has been termed the *match-mismatch* problem [28]. Programming languages, which use a highly constrained syntax, are at a relatively fixed level of abstraction (Green and Petre [29] give a working definition of abstraction). But with natural language interfaces, the grammatical possibilities are vast. Consider the variety of ways in which one can express even the relatively simple intent to set a 2-minute timer and the small number of utterances that typically produce the correct result in contemporary voice assistants. Choosing the right level of abstraction, the right level of granularity in a command, is called the *abstraction matching* problem [95].

In this work, we focus on abstraction matching, and propose an approach to improve the user experience of this problem. We acknowledge the other problems that are prior to, but distinct from, the abstraction matching problem, such as the problem of creating a well-formed intent (i.e., knowing what the system can do), and selecting tools that can solve that intent (e.g., knowing that a timer will help you avoid overcooking your egg). *Our approach does not target these issues explicitly*, although as we will see, in some instances it can have indirect benefits for these "upstream" challenges.

When conceptualized as a longer dialog, issues such as conversational breakdowns, turn-taking, and self-repair emerge [3, 59–61, 102]. We note these issues and intentionally leave them out of scope.

## 2.2 Natural language programming

Natural language (NL) has been seen as an attractive mode of programming due to its (perceived) lower learning requirements. In the 1960s, there were debates about the suitability of natural language as a programming notation [19, 32, 88]. At that time, the focus was not on translating *arbitrary* intents expressed in natural language into program code, but rather about adopting naturalistic keywords and grammars in programming languages, as in AppleScript [17], where naturalistic statements such as `set word to "Apple"` are used instead of more conventional algebraic notation (e.g., `let word = "Apple"` in JavaScript).

With developments in natural language processing (NLP), the possibility emerged of more free-form NL utterances being translated into program code (e.g., [66]). The growth of the population of non-expert computer users and end-user programmers [48, 100] gave added motivation. But NLP technology still had significant limitations, which led to unpredictable user experiences [71]. In response, researchers proposed techniques such as context-limiting, and using a reduced vocabulary (i.e., a well-specified subset of natural language in a tightly defined application context) [77]. Özcan et al. [82] survey challenges for natural language interfaces for querying data.

**Abstraction matching in large language models.** Advances in LLMs have led to the ability to solve previously intractable problems. For example, generating workflow scripts in mobile applications [2], helping non-experts design web pages by translating NL requests into CSS properties [47], summarizing code [104], and wrangling data [78].

Sarkar et al. [95] review studies of the usability of natural language programming with LLMs in particular (e.g., [43, 105, 112, 123, 124]) and articulate how LLMs have made the abstraction matching problem "fuzzy" (termed *"fuzzy abstraction matching"*): while LLMs can interpret a much wider variety of naturalistic utterances than earlier models, as a consequence, the space of utterances that may be effective at controlling the model is even more difficult and unpredictable for an end-user to learn. While a previous model might simply fail to interpret an utterance, an LLM may interpret it in a manner that is opaque and difficult for the user to generalize from.

**Technical strategies for improving code generation from LLMs.** Notable LLMs include GPT-3 [10] and LaMDA [109]. One approach to adapt a general LLM for a "downstream" task such as code generation is to *fine-tune* a *pre-trained* LLM by updating part of the model's weights. For example, Codex [16] is a fine-tuned version of GPT-3 for code-related tasks, which we use in our system.

Text generation by an LLM is seeded by a prompt: a sequence of text that somehow describes the desired output. *Prompt engineering*, the templatized design of such prompts, can improve performance on the downstream task [70], e.g., code generation. According to Liu et al. [70], approaches for prompting include appending relevant examples [41], appending a table schema [110], generating mutations of the prompt [64], summarizing complicated prompts [52], adding explanations [54], or various manipulations in the embedding space of the model [1, 46, 114]. Prompting guidelines for Codex[1] include specifying the language (e.g. `Python`), libraries (e.g. `Pandas`), exploiting comment style (e.g. `Python` doc strings and the `#` symbol), providing examples for the format or style, or organizing tasks into functions. Our approach (described in Section 4.1) involves combining the user query and data into a fragment of `Python` suitable for seeding the model.

Importantly, prompt engineering is a class of *technical* solutions for improving model performance; it positions itself as an intermediate step that augments and modifies any text written by the end-user of a natural language interface. Prompt engineering, as the term is used in the machine learning literature, is not therefore intended as a user-facing or interaction design concern.

**Design strategies for improving code generation from LLMs.** LLMs are capable of learning and generating instructions, and breaking down or decomposing a large task into smaller subtasks has consistently been found to improve the performance of LLMs [26, 37, 75, 83, 84, 115, 119, 121, 122]. Jayagopal et al. [42] examine the usability of program synthesizers, including GitHub Copilot, by novices, finding problems arising from the difficulty of task decomposition, and recommending that designers offer scaffolding for decomposition. Unlike prior work, which asks the user to decompose their request but does not consider how the user should be guided to do so, our approach can be viewed as providing an example decomposition of the generated code that the user can update and resend to the LLM as a set of instructions.

The broader area of interactive parsing frames the process of converting natural language to code as a dialogue with the user.

---

[1] https://beta.openai.com/docs/guides/code/best-practices

Previous work has explored various interaction design possibilities for this dialogue, such as generating clarification questions in natural language [125] or multiple-choice [62], helping users correct errors through natural language feedback [22, 76], formalizing user intent as tests [53], interactively "naturalizing" a commanding language [116], and guided step-by-step solution generation [101]. Weisz et al. [120] interviewed programmers about their perceptions of neural code translation tools, i.e., models that translate from one programming language into another, finding that confidence highlighting and alternative translations can improve their utility and comprehensibility. Other human-in-the-loop approaches at the level of the LLM support conversational code generation [81]. Our approach differs from these in the important respect that we focus on non-expert end-users, who have low expertise in programming (or none), and in our system users never view generated code, only the output of its execution.

## 3 GROUNDED ABSTRACTION MATCHING

The key idea of grounded abstraction matching is to systematically translate the system action (e.g., code) generated from a user query back into a naturalistic utterance that is an editable example of how to consistently invoke the same action. The resultant utterance is grounded in two ways: it has a direct, systematic correspondence to the actual code that the system generated, and it is based on an intent the user actually has.

Contrast the grounded approach with tutorial examples typically given for natural language interfaces, which are ungrounded in both senses. Take for instance an instructional example displayed by Microsoft's Cortana assistant[2]: "What's on my calendar tomorrow?". From this example, the user cannot infer a general principle for using the system in different ways. The system cannot answer "What's on *my boss's* calendar tomorrow?", or "What's on *the television* tomorrow?". This example is thus ungrounded; it bears no structured resemblance to the code generated by the system. The user experience of such systems is piecemeal; a set of disjoint commands to invoke distinct capabilities of the system, with little or no overarching grammatical structure that can be learned or used to infer the capabilities of the system. The example is also ungrounded in the user's intent: it is only useful to people who use calendars, and who often need reminding of tomorrow's schedule.

The grounded abstraction matching approach carves out a space of naturalistic utterances, which have a predictable mapping with the space of system actions. We do not force the user to only generate utterances in this space. In fact, most utterances, due to the superior performance of LLMs, have a high likelihood of being mapped to *something* in the action space. However, an utterance that matches the capabilities of the system is more likely to generate the desired result (Section 2). By recasting the user's own query in terms of this subspace, the user is exposed to grounded examples of the kind of utterances, grammar, vocabulary, and level of specificity, that is effective at generating the desired system action. Our hypothesis is that exposure to such grounded examples leads to more effective and confident use of a natural language programming interface.

There are several related problems we do not address. First, we are not directly addressing the problem of helping the user develop a well-formed intent in the first place (i.e., understanding what the system can/cannot do). In Section 6, we see that our grounded approach can help, albeit indirectly. Nor are we explicitly guiding the user to decompose a problem into smaller units. Nor are we claiming that natural language programming is more effective for spreadsheet users than alternatives (e.g., formulas, charts, pivot tables).

Nor are we attempting to *explain* the output of the model [90], although again, our results show that the grounded approach can perform some of the functions of explanation. Unlike an explanation, a grounded utterance must also be an equivalent *command* for the system. If explanation was the objective, it does not follow that language is the best form of explanation (we could have used visualization), or that an operation-by-operation restatement of the code is appropriate (we could summarize, or explain in greater detail). Both Head et al.'s Tutorons [34] and Guo's Python Tutor [31] articulate the aims of code explanation and instantiate particular solutions in particular contexts. Those systems draw on graphical and diagrammatic elements, documentation, etc., to produce an explanation.

In other words, our interface does not directly address *"this is how the system understood your query"* (input interpretation), nor *"this is what the generated code does"* (explanation), but rather *"this is how you should ask the system to do what the system thinks you just asked it to do"* (grounded abstraction matching). This differentiates our design from ostensibly similar designs, such as the query parsing (input interpretation) visualizations of Wolfram Alpha [80]. Finally, our design does not address many aspects of explanation and transparency typically considered in explainable AI research [36, 65], such as why did the model generate the output it did (and why not some other output), how did it generate it (what data and process were used), how confident is the model, how trustworthy is the prediction, how biased is the prediction, etc. The tensions between explanation and abstraction matching led to interesting design problems, which we discuss in Section 4.2.4.

## 4 SYSTEM DESIGN AND IMPLEMENTATION

We built two systems, implemented as Microsoft Excel spreadsheet add-ins.[3]

- Both systems share a common code generation, execution, and output display pipeline (Section 4.1).
- System I implements grounded abstraction matching (Section 4.2).
- System II implements an ungrounded yet viable alternative solution to the abstraction matching problem based on previously established guidelines for effectively writing queries for large language models (Section 4.3).

### 4.1 Shared code generation and output pipeline

Both systems share five technical components, summarised in Figure 2. This figure follows an example where the user has a dataset listing astronauts, the total time they spent in space, and a comma-separated list of missions they participated in. The task is to calculate each astronaut's average mission duration in hours. To begin,

---

[2]https://www.microsoft.com/en-us/cortana/

[3]https://docs.microsoft.com/en-us/office/dev/add-ins/reference/overview/excel-add-ins-reference-overview

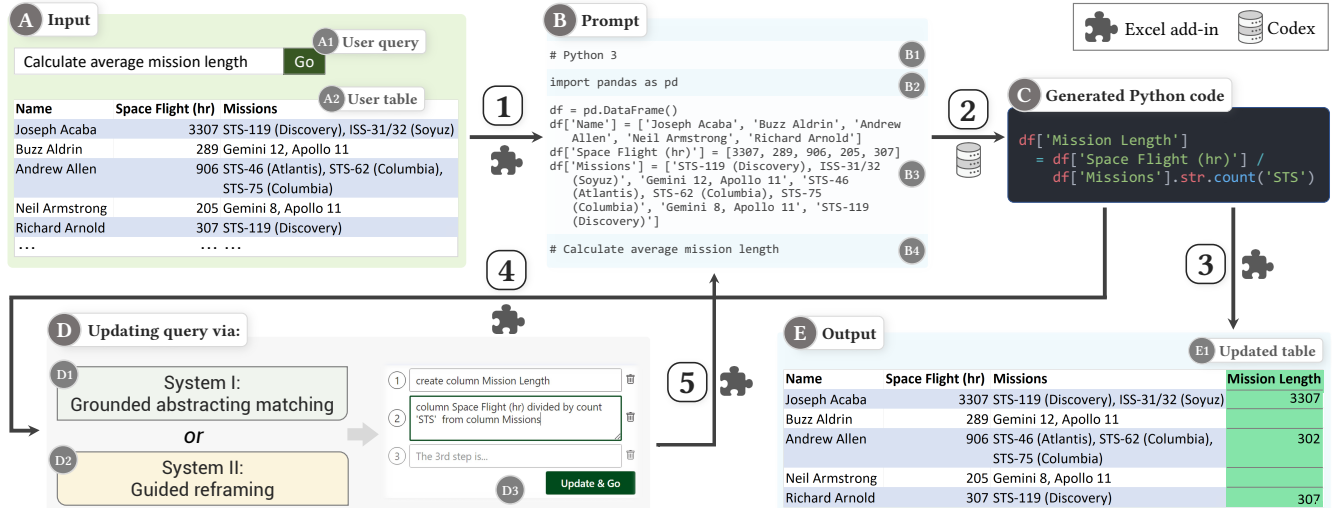

Figure 2: System architecture. (1) The user query and table are combined into a prompt. (2) The prompt is passed to Codex to generate a `Python` completion. (3) The user table is updated with extra columns or rows or a new table unless a value is returned, in which case the value is directly displayed to the user. (4) The user may update their query (in different ways depending on which system they use, i.e., (D1) or (D2)). (5) The new user query is transformed into a prompt.

the user enters the query *"calculate average mission length."* Here is how that query is processed:

1. **Conversion of user query to prompt.** Figure 2-A&B shows how the textual prompt is generated. Per best practice, we specify the target language (B1) and libraries (B2). We choose `Python` and `Pandas` both because of Codex's high performance on `Python` generation, as well as the large set of operations that are tailored to common data analysis tasks, which simplifies much data processing code into the chaining of `Pandas` API calls. The Excel table is converted to a `Pandas` dataframe (B3). The system assumes a normalized relational table. Finally we append the user query as a comment (B4). Using docstrings instead of #-style comments did not affect the performance in our setting.

2. **Code generation.** We call the OpenAI Codex API with our prompt and hyperparameters. In particular, we set the temperature to 0 (to minimize variability), and we set the stop sequence to "\n#" (i.e., at the start of a `Python` comment, as Codex tends to delimit self-contained solutions using `Python` comments). Figure 2-C shows a generated `Python` snippet.

3. **Code execution and output display.** Snippets are then run in a JavaScript web service sandbox using Pyodide[4]. This approach improves security and ensures a consistent `Python` runtime environment, since the user may not have an up-to-date version of `Python` (with appropriate libraries) installed. Figure 2-E1 shows an extra column added by writing the output from the snippet in Step 2 to the spreadsheet grid. If the completion's output is a new column or row, we append it to the user's table. If the output is a single value or a new table, we show it in a side-pane only. The user is *not* shown the `Python` snippet.

4. **User interaction.** This manifests in two options, which differ between System I and System II, as shown in Figure 2-D. In System I (D1), we generate a grounded utterance for the `Python` snippet displayed as steps, and in System II (D2), we provide a similar "step-staging" area where users can do their own problem decomposition, or provide additional hints to the system. Regardless of the option, the user can edit, add, or delete steps (D3). We expand on System I in Section 4.2 and System II in Section 4.3.

5. **Preparation of a new prompt.** When the user presses "Update & Go" (in Figure 2-D3), the steps are concatenated into a new query, loaded into the query box, with which we proceed as per Step 1.

### 4.2 System I: grounded abstraction matching

*4.2.1 Example usage scenario.* Interaction with System I is summarized in Figure 3.

Sherry is a journalist working on an article about NASA astronauts, and has gathered data about twenty astronauts in an Excel spreadsheet, including columns such as "Name", "Status" (whether an astronaut is active or retired), "Space Flight (hr)", and "Missions". "Missions" is a comma-delimited list of space missions that an astronaut has participated in, e.g., "Apollo 11", "STS-132 (Atlantis)", and "ISS-19/20 (Soyuz)".

Sherry wants to calculate an average mission length for every astronaut. For example, astronaut Joseph Acaba has a "Space Flight (hr)" of 3307 and has been on "STS-119 (Discovery), ISS-31/32 (Soyuz)", so his average mission length is (3307 ÷ 2 =) 1653.5 hours.

Sherry opens the add-in to the right (Figure 3-A), and types *"calculate average mission length"* into the query box (Figure 3-B1) and clicks "Go". This results in the generation and execution of a `Python` snippet (via the process in Section 4.1). The generated code (Figure 3-G1), which is not visible to Sherry, is:

```
df['Mission Length'] = df['Space Flight (hr)'] /
↪ df['Missions'].str.count('STS')
```
(1)

A new "Mission Length" column along with the calculated data appears in a green background (Figure 3-F1), which is also shown in the "Results" panel (Figure 3-C) in the sidebar. Sherry notices some empty cells, so she suspects there has been an error in interpreting or executing her query. She turns to the grounded utterance panel (Figure 3-D1).

---
[4]https://pyodide.org. Pyodide is a `Python` distribution for WebAssembly.

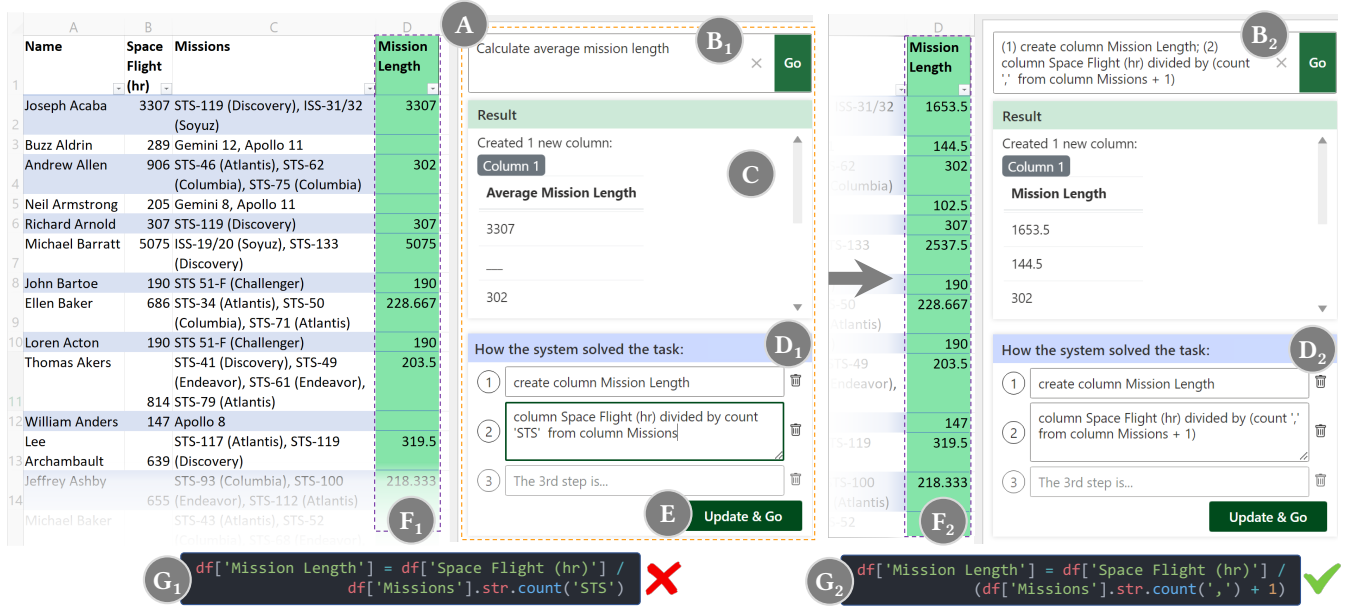

Figure 3: System I interface. Left: the user has entered a query (B1), which has been converted to code (G1) and executed by System I. The result is visible as a new column in the sheet (F1), and a grounded utterance representing the code is visible (D1). Right: the user has edited the grounded utterance via the text boxes (D1 ⇒ D2) and submitted it as a new query (B2) by clicking the "Update & Go" button (E), and new results are visible (F2).

These grounded utterances (generated according to the method explained later in Section 4.2.2) are presented as a series of editable steps (Figure 3-D1). Sherry notices in step 2 ("column Space Flight (hr) divided by count 'STS' from column Missions") the system invented a faulty heuristic for counting the number of missions (the denominator) by counting the number of occurrences of the word "STS" from the "Missions" column. Sherry observes that some rows do not contain "STS", which leads to a zero denominator and ultimately an empty cell, which also guides her to devise a correct way of calculating the mission count, namely by counting the number of commas in the "Missions" column and adding 1. After editing the second step with the new logic (Figure 3-D2), Sherry clicks the "Update & Go" button (Figure 3-E) to re-run the task, with the query being a concatenation of the updated steps (Figure 3-B2).

After reading the new result column (Figure 3-F2) and the new grounded utterances (based on the newly generated Python code, see Figure 3-G2), Sherry is convinced that the calculation is correct.

4.2.2 *Systematic grounded utterance generation.* We generate grounded utterances using program analysis. Our algorithm takes as input a Python program that is assumed to use the Pandas library, and outputs a sequence of utterances. We structure the algorithm in two parts: first, the construction of a *task-centric* program representation (TCR); and second, the generation of grounded utterances using this representation.

The TCR is designed to retain the algorithmic detail of the code whilst reducing ambiguity introduced by the concrete Python representation. Consider the code `df['Missions'].str.count('STS')`, which counts the number of occurrences of `'STS'` in the `'Missions'` column. The presence of attribute `.str` is an artifact of the Pandas library, meaningless to a user with no Python expertise. Further, consider `df['Missions'].str[0]`, which extracts the first character from the `'Missions'` column. There are two syntactic access expressions of the form `expr1[expr2]` with different meanings; the first represents column projection and the second represents string indexing. A purely syntax-driven utterance will fail to reflect these different meanings.

We construct the TCR using a type-directed translation from the Python abstract syntax tree. Types are required to resolve identifiers, such as `count`, to symbols. We can then associate utterance templates to each symbol. Types are also used to enrich utterances: the phrases *first letter* or *first word*, rather than *first element*, can be selected depending on the type of `expr1` in `expr1[0]`. At its core, the TCR is a domain-specific language for dataframes which includes row selection, column projection, column extension, variable binding, and a series of methods that operate on dataframes and columns.

We derive the utterances through a traversal of the TCR; a process we refer to as *layout*. When a TCR operation has a single *subject*, we present the operation as an instruction and layout the subject as additional instructions. For example, the code `df['Missions'].str.count('STS')` is a linear chain of operations, each with a single subject, and is therefore presented as a sequence of instructions: (1) select column "Missions", (2) calculate count "STS". In contrast, the code

`df['Space Flight (hr)'] / df['Missions'].str.count('STS')` is rooted with a binary operator, and is therefore presented as a single descriptive instruction: (1) column "Space Flight (hr)" divided by count "STS" from column "Missions". The layout algorithm can combine the instructional and descriptive styles, and therefore, the resulting utterance for example code (1) mentioned in Section 4.2.1 is: (1) create column "Mission Length", (2) column "Space Flight (hr)" divided by count "STS" from column "Missions".

Our algorithm only supports a subset of `Python` constructs and the `Pandas` library (details in Appendix B); the set is selected based on their frequency in a benchmark problem set discussed in the next section.

*4.2.3 Round-trip stability.* An important validity criterion for a grounded utterance is that running it through the code generation pipeline generates the same system action in which it is grounded. To test whether our heuristics do in fact generate utterances that have this property, we curated a benchmark dataset of questions on Stack Overflow[5] that requested help solving problems in spreadsheets. Each question and answer was distilled into an input table, a natural language query, and an expected output.

We generate a code snippet $C_1$ from the query, then generate a grounded utterance for that snippet. We then generate a new code snippet $C_2$ from the grounded utterance. We test whether $C_1$ matches $C_2$ (code generation equality) and whether executing $C_1$ and $C_2$ results in the same output (output equivalence). In practice, code generation equality is unnecessarily conservative, as Codex randomly injects statements such as "`print(df)`" that do not affect the output but will lead to trivial code inequality. Output equivalence, therefore, has more bearing on the user experience.

We calculated equivalences in both our synthetic dataset (126 queries) and the actual queries submitted by participants during our user study tasks (191 queries where a grounded utterance could be generated). In both datasets, the outputs were equivalent approximately 85% of the time, suggesting that the grounded utterances were sufficiently stable for our study of the principle of grounding. Clearly, these results are not perfect and there is room for improvement, which would be suitable for future work.

*4.2.4 Design tensions in the grounded utterance language.* In the design of the utterances, we faced a tension between using the language as an explanation, versus as a querying language. This tension manifested as many individual design trade-offs between optimizing for user understanding versus tactics for effectively guiding the model. For example, the keyword "`string`" is highly effective at specifying the type of textual content for the model, yet is meaningless to non-expert end-users. We chose "`text`" after empirically verifying that the model was performant enough to interpret this consistently. We made similar decisions between words for operations such as "`average`" (more user-friendly) and "`mean`" (aligned with the `Python` function name). An interesting and tricky case is array indexing. As `Python` arrays are zero-indexed, the model can appear inconsistent to non-programmers: asking it to produce "the first item" in an array yields `array[0]`, but asking for "the item in position 1" yields `array[1]`, which is in fact the second item in the array. Here we introduced ad-hoc intermediate heuristics that identified such references to indices and added 1 before displaying it to the user but subtracted 1 before sending it back to the system.

---

[5] https://stackoverflow.com/questions/tagged/excel-formula

Figure 4: System II interface. The portion shown replaces the section where grounded utterances are displayed in System I, otherwise the two systems are identical. The ungrounded system encourages known best practices for prompting LLMs, including (A) problem decomposition, (B) specifying output constraints, and (C) avoiding reliance on background information.

The design of these utterances is a rich space, and while we made some effort to optimize the language for our prototype, we emphasize that there is still much work to be done in exploring and articulating the various possibilities. Our solution, whilst sufficient for evaluating the concept of grounded abstraction matching, has much room for improvement, some of which we will discuss in Section 8.

## 4.3 System II: guidance reframing

We could have compared System I to one without any form of user support for learning how to use the system effectively. However, this would have been a straw man comparison, and a lost opportunity to study how different solutions to the abstraction matching problem compare. Previous work by Sarkar et al. [95] and Srinivasa Ragavan et al. [105] has already established the key problems that arise when users are given a system without support, namely: it is difficult for users to recover from errors, to develop a mental model of the capabilities of the system, and to trust the results. We were able to verify that these problems also apply to our scenario in a small formative study (Appendix C).

We thus designed an alternative system that supports users according to established best practices, but where the support is not grounded in the space of system actions or user intent. This *ungrounded* version of the system (Figure 4) interprets recently established techniques for effectively writing queries for LLMs, in the context of end-user programming in spreadsheets with LLMs. We draw upon the following practices:

(1) **Decomposing tasks.** LLMs are effective at mapping simple problems to code, e.g. those which can be expressed as a few

API or function calls, but less effective in multi-step reasoning, which requires correctly decomposing a problem [5, 75, 122]. Thus, we prompt users to "Describe the task in steps" using the same step-staging UI as System I (Figure 4-A). Users decompose the task using their own logic (contrasted with System I, where the steps can be bootstrapped through the grounded utterances).

(2) **Specifying output constraints.** Prior work [63, 75, 79, 103, 110] has found that specifying explicit textual statements of output constraints, such as the intended *shape* (a single value, a column, or a table, etc.) and *data type* (number, boolean, string, etc.) can improve performance, as the code corpora used for fine-tuning these LLMs contain comments (likely documentation strings) that specify the output type.

Thus, users are asked to specify the shape (from "a single value" to "new columns" or "new tables") and type ("number(s)", "character(s) or word(s)", or "true/false") of the system output via the controls under the prompt "I expect the result to be" (Figure 4-B).

(3) **Avoiding reliance on background information.** While there is evidence showing that code-generating LLMs can use knowledge about the world learned from natural language corpora when solving natural language tasks, it is unreliable and can lead to unanticipated side effects [75].

Thus, a "Final check" toggle asks users to confirm that their query (given the data in the spreadsheet) does not use terms and concepts that require additional knowledge to interpret (Figure 4-C). This toggle's purpose is to ensure that the user consciously designs their query to avoid requiring background information, and does not affect the prompt.

The information in these sections is concatenated and passed as a prompt to the model, as in System I.

## 5 USER STUDY: HOW DOES GROUNDING AFFECT ABSTRACTION MATCHING?

We designed a study to explore the effect of grounding on the problem of abstraction matching. Participants completed tasks modeled after real-world spreadsheet problems. We examined the queries they submitted, think-aloud data, and questionnaire responses, and address the following questions:

- In what ways does our system fail to correctly interpret the user intent?
- How do participants rewrite their queries in response to failures, to make progress on the tasks?
- How does grounding affect user perceptions of the utility of such systems, their confidence and trust in the system, and their mental models?

### 5.1 Participants

We recruited a purposive sample [24] through emails and social media. We selected for a diversity of backgrounds, including prior spreadsheet experience, formula writing experience, and programming experience. Experience was measured using a previously developed spreadsheet expertise questionnaire [93]. Participants were required to be over 18 years of age and fluent in English.

We recruited 24 participants (1 non-binary, 9 women, 14 men) across 11 industries. Fourteen participants were 25-34 years old, four aged 35-44, four aged 18-24, and two aged 45-54. Of these, half self-reported having some experience with basic spreadsheet usage while the other half reported having a lot of experience and having used at least some advanced features. Eight participants reported knowing a few basic functions (such as SUM and AVERAGE) in their spreadsheet formulas, nine reported having knowledge of advanced functions but rarely used them and preferred basic functions when writing formulas, and the remaining seven reported having built a wide variety of formulas or VBA functions. Three participants reported having little or no programming experience, seven reported having limited knowledge of programming to use it for small infrequent tasks (this is common for spreadsheet users [93], and falls within the level of expertise typical of non-expert end-user programmers [48]), nine reported being moderately experienced in programming and wrote code regularly, and the remaining five reported being highly experienced in programming.

### 5.2 Study protocol

We chose a between-subjects design, where participants were stratified and then randomly assigned to either the grounded or the ungrounded condition (the obvious learning effect of the interfaces rules out a within-subjects approach). Groups were balanced in terms of gender and prior experience. Concretely, Table 1 shows the similar gender distribution, median spreadsheet expertise, spreadsheet formula expertise, and programming expertise (as per our screening questionnaire and its integer coding scheme, taken from prior work [93]) between the two conditions.

Participants first signed a consent form and completed the demographic survey. Participants then spent 5 minutes discussing their experience with, and typical use of, spreadsheets. They were shown a brief tutorial explaining the user interface elements of the system used in their respective conditions. Participants then completed an example task, to increase their familiarity with the system and mitigate order and learning effects in the remainder of the tasks.

We designed three tasks based on real-world spreadsheet questions on Stack Overflow and other similar question-answering forums. For each task, participants were presented with a textual description of the task and a data table (of roughly 25 rows), which they read before starting the task. They were asked to think aloud [27] while completing the task. The experimenter answered questions about the task objectives and intervened to help participants recover from bugs in the system implementation, but did not intervene otherwise.

The first task was to identify how many times the city of New Orleans had won Super Bowl games (an American sporting event) given a dataset of Super Bowl records. The second was to calculate a column of average mission duration given a dataset of astronaut space flights (similar to the example in Section 4.2.1). The third was to calculate a column that checked whether a house satisfied three criteria, given a dataset of houses.

Participants were given 15 minutes per task, progressing if they finished early. Task completion was determined by the participant's own judgement, and notifying the experimenter. In rare cases, participants believed they had succeeded but had slightly misinterpreted the task (e.g., not counting a house built *in* 1970 as part of the class of houses built *after* 1970). We considered such cases as successes and did not intervene, since the participant had a well-formed intent but had simply misread the question.

Table 1: Group characteristics in experimental conditions: groups are evenly matched.

| Condition | Gender (non-binary/woman/man) | Programming expertise (median) | Spreadsheet expertise (median) | Formula expertise (median) |
|---|---|---|---|---|
| **Grounded** | 1/4/7 | 3.5 | 3.5 | 4 |
| **Ungrounded** | 0/5/7 | 4 | 3.5 | 4 |

**Avoiding priming.** We needed to avoid framing the task objectives in a way that strongly influenced user queries. While some amount of priming is inevitable, we considered multiple options that might reduce the influence of the task instructions. Presenting the task objective pictorially, by showing a screenshot of the desired output column, made it too difficult to infer the task objective. Phrasing the task differently for each participant would eliminate an overall bias, but might still bias each participant in idiosyncratic ways. Our solution was verbosity and circumlocution: we described the problem indirectly and in a long-winded manner, which we expected would encourage participants to formulate their initial query in their own terms. This strategy was effective in practice (detailed in Section 6). Details of the task descriptions and datasets are given in Appendix D.

After the tasks, participants completed the NASA TLX [33] and System Usability Scale [58] questionnaires, and engaged in a semi-structured interview probing the perceived effectiveness of the system, their practices around formulating and refining queries, debugging and verifying the system-generated results, and scenarios where they thought the system would be useful and not useful.

Each session took approximately 65 minutes, conducted via Microsoft Teams video conferencing software, with the participants remotely controlling the experimenter's computer, a designated Thinkpad laptop with Microsoft Excel and the prototype systems installed. Sessions were screen and audio recorded. Participants were compensated USD $25 or local currency equivalent. The study was approved by our institution's ethics review board.

### 5.3 Data segmentation and analysis

We transcribed think-aloud remarks and the post-study interviews for all participants. Each participant's transcript was segmented into remarks made: 1) before the tutorial task, 2) during the tutorial task, 3) during the first task, 4) the second task, 5) the third task, 6) post-study interview. The transcripts for the tasks were further segmented into "query episodes", each starting when the user submits a query and ending when the user submits the next query (or the task is complete). Thus each query episode is a single loop of the user submitting a query, observing and checking the system's output, and figuring out how to proceed if the task is still unsolved.

These segments and query episodes were augmented with telemetry, consisting of the query the user entered, whether the user submitted the query from the query box or from the step-staging area, the generated code (invisible to the user), the generated output (visible to the user), the generated grounded utterance (in the grounded condition), and any errors.

Each augmented episode was analyzed using iterative open coding [111] in accordance with Braun and Clarke's thematic analysis [9]. Think-aloud and post-experiment interview data were further analyzed to identify comments relating to the use of language, trust, and confidence in the model.

Task episodes were coded into *failure modes* and *rewriting strategies*. To generate a codebook, two researchers began by independently open-coding the same set of 52 query episodes (data from six participants). Together they generated 22 and 25 proto-codes for failure modes and rewrite strategies, which, after discussion and negotiations, led to an initial codebook of 14 failure mode codes and 17 rewrite strategies. The researchers then independently re-coded the same sample with the codebook. Manual inspection showed poor inter-rater agreement, consistent with initial open coding rounds in prior studies [11, 14, 106]. The two researchers then discussed disagreements and ambiguities and revised the codebook. The final codebook (Appendix A) consisted of 12 failure modes and 16 rewrite strategies; 14 code definitions were updated and three were merged from the initial codebook following two rounds of negotiations between researchers.

With this final codebook, the two researchers independently coded the entire set. The agreement on failure modes was 98% (researchers disagreed in 7 out of 293 query episodes), and on rewrite strategies was 57.3%. To achieve a high level of negotiated agreement [73, 87], the two researchers manually negotiated each disagreement until 100% agreement was reached on the coded set. Our use of negotiated agreement was with the intent to make quantitative comparisons between code frequencies in the grounded and ungrounded conditions. While more work may be needed to establish the reusability of the codebook as an independent analysis device, the shared agreement established is sufficient for us to draw reliable conclusions about our specific dataset. This aligns with McDonald et al.'s guidelines for reliability in CSCW and HCI research [73]. To connect our findings to our codebook, direct references to codes are presented **in bold**. Finally, the researchers grouped codes into larger themes, discussed overall findings, and selected representative query episodes and quotes.

Table 2: Number of queries issued until first solution.

| Task | **Grounded** (mean ± standard deviation) | **Ungrounded** |
|---|---|---|
| **1** | 3.17 ± 2.15 | 2.33 ± 1.31 |
| **2** | 3.50 ± 1.61 | 3.33 ± 2.75 |
| **3** | 4.50 ± 3.07 | 2.42 ± 1.80 |

## 6 RESULTS

### 6.1 Task completion and queries

All participants completed every task successfully. Participants completed the three tasks in similar amounts of time in both grounded (mean = 32 minutes 38 seconds, $\sigma$ = 7 minutes 35 seconds) and ungrounded (mean = 27 minutes 51 seconds, $\sigma$ = 7 minutes 8 seconds) conditions. The difference was not statistically significant using a t-test ($t(X) = 1.592, p = 0.1256$). We include this analysis of time taken only as an additional description of the difficulty of our tasks; owing to the variable effects of a think-aloud protocol on timing,

Table 3: Response to NASA TLX items. Format: median (mean ± standard deviation)

| Condition | Mental demand | Physical demand | Temporal demand | Performance | Effort | Frustration |
|---|---|---|---|---|---|---|
| **Grounded** | 4.0 (4.17 ± 1.46) | 0.5 (1.08 ± 1.44) | 2.0 (2.25 ± 1.59) | 8.5 (8.50 ± 0.96) | 3.5 (4.08 ± 1.80) | 1.0 (1.50 ± 1.71) |
| **Ungrounded** | 3.0 (3.75 ± 2.71) | 1.0 (1.75 ± 2.24) | 1.5 (1.58 ± 1.61) | 8.5 (8.42 ± 1.55) | 4.0 (3.92 ± 2.66) | 0.5 (1.83 ± 2.58) |

Table 4: Response to System Usability Scale items. Format: median (mean ± standard deviation)

| Question category | Statement | Grounded | Ungrounded |
|---|---|---|---|
| **Comprehensibility** | I would consider my interactions with the tool to be understandable and clear. | 2 (1.25 ± 1.36) | 2 (1.25 ± 1.30) |
| **Learnability** | I would consider it easy for me to learn how to use this tool. | 2 (1.50 ± 1.38) | 2 (1.83 ± 0.37) |
| **Enjoyability** | I enjoyed the features provided by the tool. | 2 (1.42 ± 1.38) | 2 (1.67 ± 0.62) |
| **Applicability** | Using this tool would make solving spreadsheet problems at work more efficient and effective. | 2 (1.08 ± 1.66) | 2 (1.42 ± 0.86) |
| **Recommendability** | If possible, I would recommend the tool to my friends and colleagues. | 2 (1.58 ± 1.38) | 2 (1.58 ± 0.64) |

we cannot draw conclusions on the direct effect of condition on task completion time.

Participants required similar numbers of attempts (i.e., issued similar numbers of queries) to solve each task in both grounded and ungrounded conditions. This is shown in Table 2. The difference between conditions is not statistically significant. As in the case of task 3, the grounded approach can even increase the number of queries. Our qualitative analysis will show why this is the case, and why it is not necessarily the disadvantage it may appear.

Our strategy to avoid priming the users' initial queries was to make the task description circumlocutory, as detailed in Section 5. To validate whether this strategy worked, we measure the homogeneity of the *initial* queries submitted by users for each task. The principle is that the stronger the priming effect, the more similar users' initial queries will be, having been biased towards certain words or phrases by the task description. We chose a simple metric, the Levenshtein distance [57], computed between every pair of initial queries for a given task (i.e., (24 × 23) ÷ 2 = 276 unique initial query pairs per task). The median Levenshtein distance for the tasks are 44, 61, and 88.5 respectively. These are 66%, 72%, and 86% of the median length in characters of the initial queries to each task. This can be interpreted roughly as follows: the typical pair of initial queries share only 14-34% of their textual content; they are mostly distinct and contain unique content. We consider this as having successfully avoided priming.

### 6.2 Usability and cognitive load questionnaires

Participants filled the NASA TLX [33] cognitive load scale and the System Usability Scale (SUS) [58] questionnaires after completing the tasks. SUS Likert items were integer-coded on a scale from -2 (strongly disagree) to +2 (strongly agree). The median response values are presented in Tables 3 and 4. Across both systems, participants reported a low to moderate perceived cognitive load, high perceived performance, and high usability.

The distributions of responses for each item in the TLX and SUS in the grounded and ungrounded conditions were compared using the Mann-Whitney U test [74]. Similarly, we compared the distributions of responses between participants with low and high programming experience (defined as those who responded in categories 1-3 and 4-5 respectively, to the programming experience item in our questionnaire). We compared the distributions of responses between gender groups. In all cases, we found no significant differences.

The fact that we did not detect a significant quantitative difference in cognitive load or standardized usability between the conditions is not surprising, given the high success rate in both conditions and the similar number of attempts needed for success. It speaks to the strength of the reframing principles embodied in the ungrounded condition, and shows that at least according to these metrics, the ungrounded condition is a strong, viable alternative condition and not just a straw man. However, this apparent non-result obscures significant qualitative differences between the two conditions revealed both by the strategies adopted by participants to overcome model failures, as well as in the development of their mental model and sense of agency, which we shall report in the following sections.

### 6.3 Failure modes

Our analysis revealed twelve different types of model failures, i.e., reasons why the output did not satisfy the user's intent. These could be (loosely) organized into the following four themes: technical failures, input failures, output failures, and logic failures. These were not mutually exclusive; a given query episode may result in multiple failures simultaneously. We classify these failure modes with *full visibility* of the generated code, which was not visible to the user, and as such the failure mode is not necessarily apparent to the user. The frequency of each failure mode is presented in Figure 5. There was no statistically significant difference between the failure mode distributions in the grounded and ungrounded conditions. Extended examples of each type of failure are given in Appendix E.

*6.3.1 Technical failures.* Technical failures are limitations of our code generation pipeline. In **generation failure**, no code is generated. In **execution failure**, code is generated but cannot be executed by the prototype. The user experience of each of these types of error is largely the same; the user gets no feedback except for a generic error message. Technical failures were uncommon, occurring in 55 query episodes (18.8% of the total).

*6.3.2 Input failures.* In input-related failures, the generated code operated on the wrong columns as input. This could either be a **wrong input** column selected when the correct column was explicitly specified in the query, or an instance of **soft wrong input** column, when the correct column was not specified and the model failed to infer the correct one.

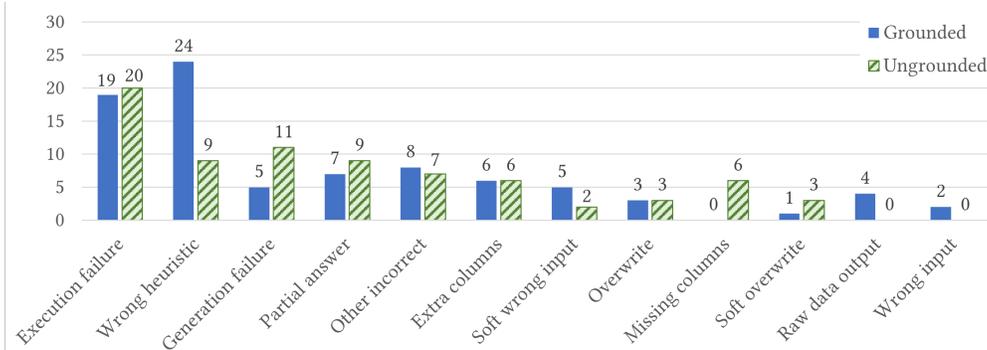

Figure 5: Relative frequency of failure modes by condition (blue: grounded, striped green: ungrounded).

*6.3.3 Output failures.* Similarly, in output-related failures, the generated code produced errors in the output format. The model might *attempt to* **overwrite** existing columns (though overwriting original data is forbidden in our current implementation) when specifically asked to generate new columns, or attempt to make a **soft overwrite** failure, when the request for a new column is implicit or ambiguous. It may generate **extra columns** which were not requested, or there may be **missing columns** which were requested.

*6.3.4 Logic failures.* Logic failures are errors in the algorithm chosen by the model to solve the problem. Arguably input and output failures, being part of the generated code, are also logic failures, but here we loosely apply the term "logic" to mean the method of computing the result. A logic failure may be a **partial answer**, which solves a portion of what the user requested, and which the user might potentially build upon to solve the remainder of the task in a subsequent query. Or, it may be **raw data output**, directly hallucinating data values to return (which may or may not be correct), rather than generating code to calculate those values. It may be the **wrong heuristic** for calculating an intermediate step when a method for computing it has not been specified or suggested by the user. Failures that do not fall into any of the above categories are coded as **other incorrect**.

## 6.4 Query rewriting strategies

How did users cope with model failures? According to *decision support* theories of interactive AI system design [50, 89], users are constantly deciding what to do next when interacting with an imperfect AI system, in particular when the model's results are unsatisfactory. In this view, the role of system design is to provide the information required to help the user decide what to do next.

We, therefore, focus on the differences between consecutive query attempts, and characterize the *rewriting strategies* that participants applied in the expectation that the changes made would solve the model failure, or at least help them make progress towards achieving their goal.

Our analysis revealed sixteen distinct rewriting strategies, which can be organized into four themes: scoping, elaboration, language restructuring, and intent shaping. These rewriting strategies were not mutually exclusive; the difference from one version of a query to the next might involve the application of multiple strategies.

*6.4.1 Scoping.* In scoping changes, participants either chose to strictly **add steps** to their previous query, asking the system to do an additional prior, intermediate, or subsequent step; or conversely, they chose to **reduce scope**, strictly removing a step that was implicitly or explicitly present in the previous query. Some examples are given in Table 5.

*6.4.2 Elaboration.* In elaboration changes, participants added more detail to their queries. They could either **elaborate how** to perform a certain calculation or computation by specifying a method. Or they could **elaborate what** they wanted in more detail without necessarily specifying how to compute it. Common ways of adding more detail about what was required were to **specify input columns** that the system should use, or **specify output type** such as new column or number, or to **name output columns**. Examples are in Table 6.

*6.4.3 Language restructuring.* Language restructuring consisted of changes to the grammar (syntax) and vocabulary used in the query. Users might break down a query into individual clauses that specify more clearly a computational step. This could have either been a **self-breakdown**, using the utterance area of the prototype to stage a sequence of steps, or in the grounded condition it could involve partial or full **reuse** (of the) **system breakdown**. In the grounded condition, participants exposed to the grounded utterances may adopt **system-like** language in subsequent queries (i.e., manually entering a query with grammar or vocabulary that mimics the grounded utterances). Or in both conditions, participants may include **code-like syntax**, such as quotes, parentheses, or even keywords from other programming languages. These are illustrated in Table 7.

Code-like syntax and system-like language were effective in improving the model output. This shows that the space of utterances that effectively controls code-generating models is not merely a subset of natural language, it is a naturalistic space that may contain elements of natural language and code. This was not lost on our participants. One participant (P1) remarked, while entering the query "rows where basement not zero" (i.e., find houses which have a basement) that it would not make sense spoken aloud: *"that one actually may be a bit less actual language, but [...] I feel like the system will understand kind of what I mean [...] I don't have to think too much about how I would actually say that in a way that would make sense [to someone else]."*

*6.4.4 Intent shaping.* The final theme consists of rewriting strategies that reflect a shift in the user's overall intent (i.e., what they want the system to do), or which help the user evaluate their intent with respect to the capabilities of the system. They may discard a strategy for solving the problem or change their interpretation of the problem and form an entirely **new intent**, or they may **start over** with the same intent but with a completely fresh expression of that intent. They may choose to partition the problem into a series of tasks, and when one task is solved, they may move on to the **next step**. Finally, they may write **testing** queries to probe the system's capabilities and improve their own understanding.

Table 5: Examples of *scoping* changes. Orange and green highlights indicate differences (removals and additions, respectively) between a query and its follow-up query (the same color scheme applies to Table 6-8 as well).

| Rewrite Strategy | Previous Query (Participant-Task) | Follow-up Query |
|---|---|---|
| **Add steps** | *Define mission_count by splitting Missions by ','* (P6-2) | ⇒ *Define mission_count by splitting Missions by ',' , then divide Space Flight by mission_count* |
| **Reduce scope** | *Create column good where year built is greater than or equal to 1970 AND squarefoot basement is not 0 and year renovated is not 0* (P9-3) | ⇒ *Create column good where year built is greater than or equal to 1970* |

Table 6: Examples of *elaboration* changes.

| Rewrite Strategy | Previous Query (Participant-Task) | Follow-up Query |
|---|---|---|
| **Elaborate how** | *Add a column of the average flight hour of each mission for each astronaut* (P5-2)<br>`df['Average Flight Hour'] = [1653.5, 190, 167, 407, 289,`<br>`↪ 302, 313, 309, 147, 1001.25, 297, 289.5, 423.5, 319.5,`<br>`↪ 205, 307, 327.5, 168.5, 343, 482.5, 2537.5, 366.5, 190]` | ⇒ *Add a column of the value of total space flight hour divided by the number of missions for each astronaut*<br>`df['Average Mission Time'] = df['Space Flight (hr)'] /`<br>`↪ df['Missions'].str.count('\(')` |
| **Elaborate what** | *how many superbowls has the city of New Orleans won* (P27-1)<br>`df[df['Host City'] == 'New Orleans'].shape[0]` | ⇒ *how many superbowls has New Orleans Saints won*<br>`df[df['Winner'] == 'New Orleans Saints'].shape[0]` |
| **Name output columns** | *Return true if year built >=1970 AND basement >0 and renovated TRUE* (P11-3) | ⇒ *Return review column where if year built >=1970 AND basement >0 AND renovated >0* |
| **Specify input columns** | *How many super bowls has New Orleans won* (P9-1)<br>`df[df['Host City'] == 'New Orleans']['Winner'].count()` | ⇒ *Select column "winner" where text includes new orleans*<br>`df[df['Winner'].str.contains('New Orleans')]` |
| **Specify output type** | *Select rows where column yr_built greater than 1970 and column yr_renovated NptEq 0 and column sqft_basement NotEq 0* (P16-3) | ⇒ *Create a column where column yr_built greater than 1970 and column yr_renovated NotEq 0 and column sqft_basement NotEq 0* |

Table 7: Examples of *language restructuring* changes.

| Rewrite Strategy | Previous Query (Participant-Task) | Follow-up Query |
|---|---|---|
| **Self breakdown** | *Count the number of rows with 'New Orleans' in the winner column* (P12-1) | ⇒ *(1) Create a new column called city that drops the last word in the winner column, (2) Count the number of rows in city that say "New Orleans"* |
| **Reuse system breakdown** | *The number of superbowls the city of New Orleans has won* (P5-1)<br>`df[df['Host City'] == 'New Orleans']['Winner'].count()`<br>(1) select rows where column Host City is New Orleans,<br>(2) select column Winner,<br>(3) count. | ⇒ *(1) select rows where column Winner is New Orleans Saints, (2) count*<br>`df[df['Winner'] == 'New Orleans Saints'].count()`<br>(1) select rows where column Winner is New Orleans Saints,<br>(2) count. |
| **System-like** | *how many super bowls has New Orleans won* (P9-1) | ⇒ *select column winner where text includes new orleans* |
| **Code-like syntax** | *Create a column that shows the value of Space Flight (hr) divided by the number of items of the Missions column* (P26-2) | ⇒ *(1) create column Space Flight (hr) per Mission, (2) column Space Flight (hr) divided by (count , from column Missions + 1)* |

Table 8: Examples of *intent shaping* changes.

| Rewrite Strategy | Previous Query (Participant-Task) | Follow-up Query |
|---|---|---|
| **New intent** | *Select rows where basement > 0 and yr_built >= 1970 and yr_renovated > 0* (P21-3) | ⇒ *Create a column called consider? where the value is true if basement > 0 and yr_built >= 1970 and yr_renovated > 0* |
| **Start over** | *Use winner column to subtract the winner team column to get the winner city* (P22-1) | ⇒ *Create a new winner city that remove the winner team from the winner column* |
| **Next step** | *Count the number of mission in column I delimited by comma* (P24-2) | ⇒ *For each name calculate the hours of space flight divided by the mission count* |
| **Testing** | *Create a new column that counts how many strings are separated by ',' in the "Missions" column. Create another column that calculates "Space Flight (hr)" divided by "Missions_Count"* (P25-2) | ⇒ *new column: # of strings separated by "," in "Missions". Create another column that calculates "Space Flight (hr)" divided by "Missions_Count"* |

Examples are in Table 8. Rewrite strategies that did not fit any of the above categories were marked as **other rephrase**.

6.4.5 *Rewriting strategies differ between grounded and ungrounded conditions.* The frequency of each rewrite strategy code is presented in Figure 6. There are some differences between the frequencies in the grounded and ungrounded conditions. One trivial difference is that the grounded condition enables the **reuse system breakdown** and **system-like** language restructuring strategies. These are naturally absent from the ungrounded condition.

| Condition | Task | Add steps | Reduce scope | Elaborate how | Elaborate what | Specify input columns | Specify output type | Name output columns | Self-breakdown | Reuse system breakdown | System-like | Code-like syntax | New intent | Start over | Next step | Testing | Other rephrase |
|---|---|---|---|---|---|---|---|---|---|---|---|---|---|---|---|---|---|
| Grounded | 1 | 2 | 0 | 10 | 3 | 20 | 7 | 3 | 0 | 8 | 4 | 7 | 21 | 5 | 3 | 8 | 4 |
| Ungrounded | | 1 | 0 | 7 | 4 | 18 | 12 | 4 | 2 | 0 | 0 | 6 | 20 | 2 | 2 | 6 | 5 |
| Grounded | 2 | 2 | 2 | 22 | 0 | 16 | 14 | 9 | 0 | 12 | 0 | 7 | 22 | 2 | 5 | 8 | 2 |
| Ungrounded | | 4 | 1 | 26 | 1 | 16 | 17 | 14 | 6 | 0 | 0 | 14 | 17 | 7 | 3 | 13 | 12 |
| Grounded | 3 | 3 | 3 | 15 | 1 | 16 | 12 | 9 | 1 | 5 | 3 | 9 | 25 | 1 | 11 | 8 | 14 |
| Ungrounded | | 0 | 0 | 15 | 0 | 18 | 14 | 14 | 5 | 0 | 0 | 19 | 17 | 3 | 2 | 11 | 12 |
| **Total over tasks** | | | | | | | | | | | | | | | | | |
| Grounded | | 7 | 5 | 47 | 4 | 52 | 33 | 21 | 1 | 25 | 7 | 23 | 68 | 8 | 19 | 24 | 20 |
| Ungrounded | | 5 | 1 | 48 | 5 | 52 | 43 | 32 | 13 | 0 | 0 | 39 | 54 | 12 | 7 | 30 | 29 |
| | | Scoping | | Elaboration | | | | | Language restructuring | | | | Intent shaping | | | | |

Figure 6: Frequency of rewrite strategies per task and overall.

Not including the trivially different codes, the overall distribution of code frequencies in the grounded condition is different from the ungrounded condition with statistical significance ($\chi^2(13) = 70.3, p = 7.1 \cdot 10^{-7}$). In the following paragraphs, we will focus on differences observed in the frequencies of the **next step**, **reduce scope**, and **start over** strategies, which, combined with qualitative analysis of the think-aloud data, explain some of the key advantages of the grounded strategy.

We see greater use of the **next step** strategy in the grounded condition and particularly in task 3. This is because participants were able to recognize a partial answer much more effectively using the grounded utterance, and they were more likely to choose to build upon the partial answer in later steps. In task 3 (finding houses to satisfy 3 criteria), the model would often give a partial answer by only satisfying 1 or 2 criteria, or by giving 3 separate columns and not combining them. When faced with this scenario, participants in the grounded condition were more likely to recognize this as a partial answer due to the grounded utterance making it explicit what the system had done. In contrast, in the ungrounded condition, participants were more likely to disregard such apparently partial answers and attempt a **self-breakdown** or introduce clarity via **code-like** syntax.

The **reduce scope** strategy, though rare in absolute terms (probably because the tasks were of small scope to begin with), occurred relatively much more often in the grounded condition. The grounded language, as hypothesized, served as a reference point for the granularity of command that is achievable by the system, i.e., the complexity that can be expressed in a typical `Python` statement.

The **start over** strategy is particularly interesting, because it was more common in the grounded condition in task 1, but in the ungrounded condition in tasks 2 and 3. This is because the decision to start over is a cost-benefit tradeoff: the cost of starting over is high in longer and more complex tasks (such as tasks 2 and 3) but is lower in simple tasks (such as task 1). The grounded utterances helped participants evaluate this tradeoff more effectively; in task 1, it was easier for participants to understand the system failure and start over, but in tasks 2 and 3, it was easier for participants to understand the system failure and adapt their query, thus *avoiding* the need to start over. Conversely, without the feedback of the grounded utterances, participants in task 1 were stuck with ineffective approaches longer than necessary, and in tasks 2 and 3 were likely to abandon results that were partially correct because they could not recognize them as such.

## 6.5 Perceived utility of grounded utterances, and their effects on trust and mental models

From think-aloud comments and semi-structured interviews, we found that grounded utterances facilitated explanation and debugging, increased users' trust and confidence, and shaped users' mental models of system capabilities.

*6.5.1 Grounded utterances facilitated explanation and debugging.* Despite the fact that explanation was not our main objective, grounded utterances gave participants a way of comprehending the system's behavior, by manifesting the system's problem-solving logic and key information. P3 remarked that *"the breakdown would help me just be able to check that what I've typed in did actually make sense to the system and that it did actually do what I was hoping."* P1 explained that *"[the breakdown] helps me understand what's going on, and therefore whether the results is going to be accurate or not,"* and participants imagined it being increasingly valuable with larger tables, where it is infeasible to manually check all rows and columns.

The presentation of the grounded utterances as isolated steps, each with reduced scope and standardized explanation language, exposed multiple entry points for users to identify and repair bugs. In general, participants thought that the grounded utterances *"made it easy to check your work"* (P4), highly *"programmable"* (P5), and *"providing opportunities for you to modify and iterate over it"* (P26). Grounding for debugging was especially useful for participants with very little or no programming experience. For example, P24 reflected that *"in the New Orleans [task], immediately, I saw that it picked from the 'Host City', and I knew this is why you're giving me [an incorrect answer] and then I had to redirect it to go look at the 'Winner' column."* Even a few exposures to the grounded utterances influenced participants' rewrite strategies (discussed in Section 6.4), as opposed to having to employ guesswork to debug in the ungrounded case.

*6.5.2 Grounded utterances increased users' trust and confidence.* In addition to the ease of comprehension and debugging discussed in Section 6.5.1, participants commented on the fact that the grounded utterances aligned with their *"intuitions on how to solve a problem"* (P22). For example, P13, who reported having little prior programming experience, recalled that *"the step-by-step approach felt natural and very much mirrored what I usually do, which is to create these temporary columns as I go along."* Meanwhile, participants with programming experience also thought the grounded utterances felt familiar, e.g., *"it's like a more natural language version of SQL"* (P16).

Participants, especially non-programmers, also felt that having access to the grounded utterances made it easier for them to trust the output, for example, *"the more you use [grounded utterances], the more confident you'll get to the values, and you get less worried"* (P24). Furthermore, despite having some prior programming experience, P5 imagined a future in which she would *"rely more and more on the add-in to do my work."* In general, the informative signals from the grounded utterances contributed to comprehension, debugging, and a sense of intuitiveness and familiarity, which in turn contributed to participants' trust in the system's behavior and confidence in their ability to steer the system toward a desirable result.

*6.5.3 Grounded utterances shaped users' mental models of system capabilities.* Interacting with the grounded utterances enabled participants to develop mental models of the system's capabilities and limitations. To some (8/12), being specific about what is being asked for helped with successful LLM generations, reflected in rewrite strategies such as **elaborate what**, **name output columns**, **specify input columns**, and **specify output type**. P6 explained his choice of mentioning the column name in his queries for task 3: *"I went to the exact column name because that's a useful reference [for the system], and I don't super trust the system to be able to semantically determined that a renovation is connected to 'year_renovated'."*

Participants picked up vocabularies and styles of utterance from the grounded utterances, which would reliably get the system to work according to their intent (7/12). This was particularly helpful for non-programmers, for example, P13 recalled that through interacting with the breakdowns, *"I can see what it's working with, what words and language and vocabulary it's working with, and then I can kind of shift my understanding of what it wants me to say or what it understands the best."* Meanwhile, some participants even attempted to map the breakdown texts to programming or script languages that they were already familiar with, for example, *"when I read and [subsequently] wrote words like 'select', I was very much thinking about SQL"* (P9) and *"this [grounded utterance] would be very similar to how I would tackle it in R."* Participants understood such styles of utterance to be *"reusable and transferable"* (P16) across different tasks, i.e., they could generalize from grounded utterances to form a predictable notion of a commanding language.

Much like re-purposing spreadsheet formulas [44], 4/12 participants saw the value of keeping track of the grounded utterances of successful LLM generations so that they can act as *"informal documentation"* (P26) for future selves and collaborators to better understand the original intent and the system's calculation process.

## 7 LIMITATIONS

We only generate grounded utterances for a subset of the Pandas/Python APIs, selected by frequency in our benchmark dataset. This was sufficient for our study: an explanation failure occurred only 13 times out of 159 queries in the grounded condition, on average 1.08 times per user. We do not claim that our algorithm is the most effective, and future work could explore alternative ways of producing grounded utterances, such as leveraging the LLM itself [45, 72]. Our system is limited in assuming a single well-defined relational data table. Future work may investigate handling multiple tables [126] or automatic table detection [20].

We chose the reframing principles from previous work as a viable alternative to our grounded approach, but there are other options, such as examples or tutorials. We chose an interface that does not require the users to encounter any information over and above the text in the user interface, on the basis that in a realistic commercial spreadsheet feature, the user is not interested in learning and prefers to develop skills through usage (the "paradox of the active user" [12]). However, this is not a concern for every natural language interface, and future work should explore alternatives.

A decision support loop restricted to rewriting queries is not the only possibility; there is a large design space which may increase the decision surface to the user, including data-oriented ("observation-level") interactions [23, 98] (such as indicating incorrect outputs [91, 96], or manually giving correct examples [30]), or access to adjust parameters of the model [51, 92] such as temperature. In our prototype, the usability of naturalistic utterances as a method for controlling a large language model is the central concern. Thus rewriting and resubmitting a query is the only available response to a model failure in our study. This allowed us to inspect the use of language closely, but future work may explore the interaction of language with these alternatives.

The data tables in the study consisted of 20-30 rows. In practice, spreadsheet data analysis can contain much fewer or more rows. With large datasets, it becomes impossible to verify the model's output for each row. A large dataset would have unnecessarily increased the complexity of the debugging task, which would confound our investigation of grounding, and which we leave for future work.

A lab study cannot capture long-term effects, which become apparent with days, weeks, or months of use. What appears to be a clear advantage of one approach over another may erode with user practice and learning. What appears to be an insignificant difference may compound over time to create a marked gap. To longitudinally validate our findings, future work may conduct diary studies [86] or experience sampling [18].

Many of our participants had prior expertise in formulas and programming. It is likely that most users do *not* regularly use formulas, based on corpus estimates [97]. Our sample reflects the important segment of spreadsheet users who do write formulas, and because the interaction design of our system avoids any direct inspection or authoring of code, we have reason to believe that some of our findings might generalize to non-programmers as well. Future work may explore specifically how grounded abstraction matching might help users without any prior exposure to formulas or programming.

# 8 DISCUSSION

## 8.1 Comparison with related work

Our findings throw additional light on prior work. Setlur and Tory [102] evaluated the interaction design of a chatbot for data analytics. We further explore how a non-expert end-user, working with a natural language interface in a data analytics setting, can be guided to query the system more effectively. Our rewrite strategies expand upon their classification of follow-up utterances into simplification and clarification categories, and we also expand their analysis of failure cases. They observed the need to *"support query expressibility"*, which we directly address with grounded utterances. Our findings support their observation that *"predictability [...] for handling different types of analytical questions further enhanced people's trust"*.

Our findings also support certain conclusions of Jayagopal et al. [42], in particular, their analysis of Lau's design guidelines for PBD systems [56] in the context of LLM-based code generation. Our findings support guidelines 2: *"Make it easy to correct the system"*, 3: *"Encourage trust by presenting a model users can understand"*: as we observed in Section 6.5. Our use case shows a new facet of guideline 5: *"Consider the perceived value of automation"*, which considers the cost-benefit trade-off of the system as a whole. We found that even within each episode of use, there are smaller cost-benefit trade-offs, such as whether to continue refining a particular query or start over.

Ragavan et al.'s study of natural language formulas in the spreadsheet grid [105] is a close precedent to our work, as it shares the application domain (data analysis in spreadsheets), target end-user, and natural language interface. We find significant commonalities. For instance, our participants also clearly stated the ease of use of natural language as an advantage compared to spreadsheet formulas or programming language. Their analysis of failure cases identified causes such as using incorrect or ambiguous words for concepts, or phrases requiring background knowledge to understand: our data shows how both grounding and established reframing principles can help users avoid, detect, and recover from such issues. We also extend their work. They acknowledge the limitations of their system that *"In cases of [...] errors, a user needs to understand how the intelligence apparatus has interpreted their utterance, and how they can fix it"*, and that *"users need a way to [...] provide an alternative phrasing for the task, when the intent is misinterpreted"*; our findings show what happens when these are provided, namely: that the ability to rewrite the query manifests in several rewriting behaviors, and an intelligibility mechanism (such as grounded utterances) can shape these behaviors.

## 8.2 The tutorial value of grounded utterances

A recurring theme from participant feedback was that grounded utterances could not only be directly reused on future tasks with the same goal, but also serve as examples of canonical ways of expressing intent, much like example code for calling an API in documentation [68]. Some participants who are not expert spreadsheet or formula users felt it much more straightforward to *"work with natural language than trying to recall or search for that exact formula for formula combination"* to perform data analysis tasks (P5), observing that *"even if I can't figure out what Excel functions to use, at least now I know how to sort of just say what I want in English and make sure to hit those keywords like 'create a column', 'split by', 'select rows where '"* (P26).

The grounded utterances can serve as an *"educational tool"* (P22) for learning logical thinking and problem decomposition skills. This has two advantages: 1) the step-by-step utterances are grounded in a sequence of API calls in the generated code that are by themselves primitive building blocks for computation; 2) LLMs are trained on large corpora of data and are usually able to "translate" a reasonably specified initial intent by a user into some logical code, as documented in Section 2 and evidenced by our system.

The way that an LLM solves a task may be suboptimal (e.g., being time and space inefficient, reflecting poor practices in coding, etc.), and could lead to users being misguided downstream. However, in our study, grounded utterances were presented as a series of editable steps exposing the abstractions of the LLM-generated code, which enabled them to reason about model failures (Section 6.3) and edit the utterances to fix model mistakes (Section 6.4). Helping learners build mental models for chained functions is a topic of interest in educational tools, and future work may explore alternative visualizations, such as those in DS.js [128], or Pandas Tutor [55].

## 8.3 Genres of naturalistic commanding

The notional language of queries that users form is surprisingly easy to influence. We observed that even a single exposure to a grounded utterance can change the grammar and vocabulary of subsequent queries. Participants looked for cues in the task question (although they were largely thwarted by our strategy of circumlocution), in the language of the interface, and in the language used by the experimenter. Moreover, they draw upon their previous experiences of querying search engines, their expertise (if available) in formula programming and other programming languages, and syntax from algebraic notation. The process by which users form a coherent query language from these disparate, fragmented influences resembles the process of creolization as articulated by Bickerton [6]. Language bioprogramming theory [7], which assumes an innate capacity for grammar, may be a partial explanation for how users transform their priming into a structured language.

However, the similarity is superficial: creolization typically occurs at a generational timescale, whereas users of these systems resolve their influences into a querying style within minutes. Unlike a creole, the user is the only speaker of their particular query language. We observed substantial individual differences in querying styles (without substantial differences in success rate). Thus each individual appears to develop their own "speech genres", per Bakhtin [4]. This flexibility can be an advantage, but also a disadvantage when it comes to collaboration and communication; as several participants noted, the query forms a part of the documentation of the spreadsheet to be passed to collaborators for comprehension. But this is another manifestation of the common fallacy committed by spreadsheet authors [106]: that what is intelligible to them is therefore also intelligible to a different reader.

Nonetheless, it is possible that through collaboration, certain norms, standards, and best practices might emerge, which will lead groups of individuals to a shared style of naturalistic commanding. However, these will be slightly different for every system and its method for generating grounded utterances, which may be vendor-specific. The user might learn one style of naturalistic commanding for a spreadsheet software made by one company, but another for database software made by another company. This confusion of "dialects," if we may call them such, is a potential interaction design challenge for the future of these interfaces.

### 8.4 Applications of grounded abstraction matching

Grounded abstraction matching is a general technique for familiarizing users with the space of utterances effective for commanding any particular language model. Here we have applied it to a system where the space of system actions is short Python data analysis programs using the Pandas library.

Even within spreadsheets, there are other applications. For example, a query seeking a particular presentation style in a spreadsheet, such as coloring alternate rows in red, could be solved using spreadsheet presentation scripting APIs, and grounded utterances could help the user learn how to express their preferred styles. Similarly, a query such as "generate a scatterplot for each data series and mark all negative $x$ values on the plots in red", could result in the generation of a Vega-Lite [99] visualization, and grounded utterances could be used to allow the user to learn the space of effective utterances and grammar of visualizations.

Outside of spreadsheets, grounded utterances could be integrated into the feedback loop for commercial voice assistants, chatbots, or software that rely on naturalistic queries, such as search engines.

While most LLM applications currently use naturalistic utterances as their input space, there may be applications where the input space is not a linguistic notation. For example, as part of an accessibility device, an LLM may be used to "translate" from a space of non-lingual speech sounds, or a space of body movements, or gestures, into a space of system actions. Here again, grounded "utterances" can orient the user to an effective use of the LLM.

### 8.5 Continued applicability of grounded abstraction matching as LLMs evolve

At the time of writing, LLMs are improving at a rapid pace, with yearly dataset and parameter scaling consistently achieving emergent properties [118]. Based on current research trajectories, it is not unreasonable to forecast that LLMs will improve in two directions: they will be more frequently able to offer "zero-shot" solutions (i.e., without bespoke fine-tuning) in current scenarios, and they will be able to tackle new scenarios.

LLMs can now provide support in many scenarios that were previously intractable [8]. The technical capability has gone from not being able to offer any assistance to being able to offer a wide variety of compelling and viable, yet poorly-understood and unpredictable assistance. The key challenge for interaction design, therefore, is finding appropriate application domains, and helping users make the best use of LLMs while accounting for their limitations. The specific problem of abstraction matching, along with the related problems of explanation and trust, are unlikely to disappear with better performance; rather as LLMs become more performant they are likely to be applied in increasingly complex and high-risk applications. The interaction design principle of grounded abstraction matching is generic, yet prescriptive enough to be helpful to system designers in many of these situations.

### 8.6 Implications for design

Our work is an early exploration of grounded utterances as a solution to the abstraction matching problem. It is not straightforward nor entirely appropriate to directly prescribe implications for design [21, 107]. Nonetheless, some of our findings may help design.

Abstraction matching becomes a serious challenge when the space of system actions is large (as with Python code). Fuzzy abstraction matching becomes a serious challenge when the language understanding model is highly performant but still highly unpredictable (as is Codex). In such situations, grounded abstraction matching can be a systematic and effective way of getting users familiar with a naturalistic commanding language.

Designers should be aware of language priming cues in the user interface and also other cues that the user base is exposed to, e.g., from other software, programming languages, search engines, etc. that they commonly use. These priming cues can be built upon, e.g., by borrowing naturalistic keywords from languages such as SQL. Conversely, a mismatch between environmental cues and the language needed to effectively command your system can interfere with the development of mental models.

Grounded utterances have the potential to serve multiple functions simultaneously: as tutorial examples for the user, as input interpretation, and as system behavior explanations. Ideally, a natural language interface would treat each of these separately, but in commercial tools such as spreadsheets, the user will not expect to attend to many different categories of feedback just to get their data analysis done, and could get overwhelmed. Thus a consideration for feedback from any natural language interfaces ought to be whether it can serve "double duty" as grounded examples (testable through round-trip experiments) as well as explanations.

## 9 CONCLUSION

Abstraction matching, selecting a natural language utterance that is likely to be understood correctly by the system, is a core problem facing users of almost all natural language interfaces. We propose *grounded abstraction matching*, in which grounded examples of such effective utterances are systematically generated and shown to the user. We present a concrete instantiation of grounded abstraction matching in a system that helps non-expert end-user programmers perform data analysis in spreadsheets.

In a study comparing this approach to an ungrounded alternative, we find that the grounded approach has many positive effects on the strategies available to end-users to cope with model failures. We find that over time, exposure to grounded examples leads to a more consistent mental model, greater confidence, and perception of trust in the system. There are many avenues for future work, including studying the effect of such grounding in dialogue, in other contexts besides spreadsheets, studying the effects over longitudinal usage, and how different "dialects" of naturalistic language that arise due to differences between users and between systems might interact.

### ACKNOWLEDGMENTS
We would like to thank our study participants for their kind participation. We sincerely thank Sruti Srinivasa Ragavan, Ian Drosos, and Sherry Tongshuang Wu for their insightful feedback and constant support.

# A CODES AND DESCRIPTIONS
## A.1 Failure modes

Table 9: Failure modes and their descriptions

| Failure mode | Description |
|---|---|
| Generation failure | No completion from Codex. |
| Execution failure | Code could not be executed. |
| Output type failure | Our prototype could not parse the output type (e.g., the model generate a column, dataframe, or value, but the format cannot be displayed in the spreadsheet). |
| Wrong input | Model generates code that calculates answer using wrong columns (even if user instruction does not specify which input columns to use, but researcher can infer it from context). |
| Soft wrong input | Like wrong input, but it is difficult for researcher to infer correct columns from user query. |
| Overwrite | Model attempts to overwrite existing dataframe column when specifically asked to create a column. |
| Soft overwrite | Model attempts to overwrite existing dataframe column, but not asked by user to specifically create a new column. |
| Other incorrect | Model generates code that is incorrect, but none of the other more specific codes applies. |
| N/A | Model generated code matches user intent (success, either intermediate or overall). Or user playing around e.g., "hi." |
| Raw data output | Model hallucinates data output (array of values) directly, not a computation. Considered a failure regardless of whether values are correct or not, because it is not reproducible on other data sets. |
| Missing columns | Model generates wrong number of output columns, fewer than the user asked. |
| Extra columns | Model generates wrong number of output columns, more than the user asked. |
| Wrong heuristic | Model invents a plausible way of calculating a quantity which is asked for but the method to calculate it is unspecified, but the method is wrong. |
| Partial answer | Model generates an intermediate step that the user can build on to complete task, OR model only correctly answers a clearly delineated subpart (e.g., in the breakdown style, or separated by AND or comma) of the user query. |

## A.2 Rewrite strategies

Table 10: Rewrite strategies and their descriptions

| Rewrite strategy | Description |
|---|---|
| Add steps | User is asking the model to do strictly more than in the previous query, e.g., adding a new follow-on step, or a new intermediate column. |
| Code-like syntax | User borrows syntax from coding languages, e.g., "()" to group statements, use of quotes, use of "if/then/else", use of symbols like ">=", use of SQL-like keywords. But not specifically the style of the system-generated breakdowns. |
| Next step | Marks when the user moves on to a new query intent that is a following step to their overall task solution. |
| New intent | User decides to try asking the system a different thing (i.e., different code is needed to solve the new intent - not just a rephrasing to ask the system to do the same thing). |
| Name output columns | Specify a name for the desired output column |
| Specify output type | Add information about the number of columns, type of columns, or type of values expected. (Not if already accounted for by "name output columns"). |
| Specify input columns | Add information about which columns the code should use to calculate the answer name output columns Specify a name for the desired output column. |
| Reuse system breakdown | Partial or full reuse of system-generated breakdown. |
| Self-breakdown | Expresses query in clearly delineated steps, but text is not system-generated. |
| Elaborate how | Give more detailed logic or computation steps for something that user was already asking for previously (e.g., *"calculate average"* -> *"calculate average by dividing x by y"*). |
| Elaborate what | More detailed description of the required output (but not including new details of how to compute it). E.g. *"count"* -> *"count the rows"*. |
| Testing | User experimenting, playing, rolling back etc. |
| Reduce scope | User asks the system to do less, or a subset, of previous query. |
| System-like | User borrows style/syntax/vocabulary from the system-generated explanations, but not a reuse (i.e. the user types it manually). |
| Start over | User appears to abandon previous query text entirely; new query text contains little or no character sequences in common with previous query (*and* the intent is the same, i.e. not new intent). |
| Other rephrase | User rewrites part of the query, but none of the other codes applies to that specific part. (other parts may be rewritten in a different manner which we can code). |

## B SYSTEM IMPLEMENTATION DETAILS

### B.1 Further details of grounded utterance generation

The implementation of utterance generation is specific to a programming language or API, in this case Pandas, but the design of the algorithm is generic. The algorithm is split into three components:

(1) Translation of python code into the task-centric representation (TCR).
(2) Translation of TCR into an explanation representation (ER), a tree of natural language fragments, preserving hierarchical structure.
(3) The layout algorithm that converts an ER tree into an utterance. The algorithm is parameterised such that it can generate a single sentence or a sequence of utterances.

As the algorithm consumes and produces different trees, starting with a Python abstract syntax tree, the algorithm works with single or multi-line programs. The features that determine what is supported by the algorithm are the language constructs and API methods used, rather than how the code is written.

For this work we focus on defining the algorithm over a subset of Pandas that covers basic data analysis and wrangling tasks. This includes:

- Dataframe indexing operations such as `loc` and `iloc`, supporting patterns that includes element access, slice access, and masking with boolean-valued series.
- Vectorised operators applied to both dataframes and series.
- Variable declarations bound to pandas expression.
- Dataframe methods such as `groupby`, `size`, and `transpose`.
- Aggregation methods such as `sum`, `min`, and `max`.
- Series methods such as `idxmax`, `sum`, and `mean`.
- Series string methods such as `split`, `replace`, `lower`, and `strip`.
- Series date and time methods such as `ceil` and `year`.

Notable Python features not currently supported include function declarations, list comprehensions, and control flow.

To improve the generated utterances we add special handling for certain patterns. Examples include:

- Labels for tuple types. Certain Pandas operations return tuples with a defined meaning, such as `.shape` which returns the dimensions of a dataframe. Our algorithm understands that `.shape` returns a tuple type, and additionally, we allow the type to label each element of the tuple, such as *rows* and *columns*. When generating an utterance for a subscript operation, such as `e1[e2]`, if `e1` is inferred to have a labeled tuple type and `e2` is a known constant, we use the label in the utterance rather than a generic *"element of"* snippet.
- Strings. When accessing elements of a string column which contains either a single string or a list of strings, we use the terms *"character"* or *"word"* instead of *"element"*.
- Array indexing. As Python arrays are zero-indexed, we add 1 to indices before displaying to the user (e.g., `array[1]` is rendered as "element 2 of the array"), and subtract 1 from user queries, detecting such indices using templates, before sending them to the system.

### B.2 Formal description of round-trip benchmark

The benchmark simulates the following steps for each input table $I$ and natural language utterance $Q$:

(1) The user selects table $I$, types $Q$ into the query box, and clicks "Go".
(2) The system generates code $C_1$ and output $O_1$ (which can be a single value, new column(s), or new table(s)).
(3) If possible, the system generates a grounded utterance $G_1$ from $C_1$; otherwise the interaction ends.
(4) The user clicks "Update & Go" without editing $G_1$.
(5) The system generates code $C_2$ and output $O_2$.

To measure the equality of results, the benchmark calculates two metrics:

- **Code generation equality**: how frequently $C_2 = C_1$ given that $G_1$ can be generated.
- **Output equivalence**: how frequently $O_2 = O_1$ (modulo column names) given that $G_1$ can be generated.

We calculated equivalences in both our synthetic dataset and the actual queries submitted by participants during our user study tasks (Section 5). Table 11 summarizes the results.

Table 11: Results of the round-trip stability experiments on both the Stack Overflow dataset and actual participants' queries from the study. $N$ is the number of data points that successfully passed the aforementioned 5 simulated interactions. Both equality measurements are presented as the percentage of data points where equality was reached.

| Dataset | $N$ | Code generation equality | Output equivalence |
|---|---|---|---|
| **Stack Overflow dataset** | 126 | 58.7% | 85.7% |
| **Study participant dataset** | 191 | 72.8% | 84.8% |

## B.3 Evaluating LLM performance

In order to assess a LLM's performance for code generation, the research community has developed a wide range of benchmarks and quantitative metrics. Primarily there are two types of evaluations that are conducted: output matching and code equivalence. For output matching, we mark a generation as correct if it passes all I/O examples or tests (as the generation process is not deterministic we use *pass@k* to estimate the chance that we obtain a correct generation [16]). For code equivalence we use code similarity metrics [25] where we mark a generation as correct if it is similar enough to the expected code snippet. The most used benchmarks target Python generation (e.g. APPS [35], HumanEval [16]) and SQL generation (e.g. Spider [127]), but recently benchmarks for cross-lingual generations have emerged (e.g. [13], XLCosT [129]) as well as benchmarks with cross-lingual asks (e.g. MCoNaLa [117]). In our work we evaluate over a new dataset of problems extracted from StackOverflow under the tag Excel-Formulas, where the users believe their ask can be solved by an Excel formula. We use output matching (pass@k) as our metric for correctness.

## C FORMATIVE USABILITY STUDY

We conducted an informal, exploratory study to gain an initial understanding of the challenges users might face when interacting with an LLM for data analysis in spreadsheets, and to check whether and to what extent the abstraction matching challenges observed by Sarkar et al. [95] and Srinivasa Ragavan et al. [105] were present in our application.

### C.1 Prototype

The prototype used in this formative study uses the code generation pipeline described in Section 4.1, specifically steps 1-3. These steps take a user query, generate executable Python code using Codex, run the code, and insert the results into the spreadsheet. However, the user interface of this prototype was minimal. Users could only enter queries using a query box (similar to Figure 3-B1/B2) and subsequently view the execution results in the spreadsheet (similar to Figure 3-F1/F2). There was no guidance of any kind for writing effective queries. This user experience resembles commercially available spreadsheet querying features, such as the "Analyze Data" feature in Microsoft Excel,[6] or the "Explore" feature of Google Sheets.[7]

### C.2 Method

We recruited 5 participants (FP1 to FP5) through UserTesting.com. Participants were end-user programmers who all self-reported writing formulas in spreadsheets for work. One of the participants had some prior programming experience using C++, while the rest had no programming experience. There were no participants in common between this sample, and the sample used in the main study. While we do not claim that this sample is representative of all users, the study sessions were informative and helped motivate the idea of grounded abstraction matching, the development of both System I and System II, and the design of the formal study in Section 5.

We began by asking participants about their experience with, and typical use of spreadsheets. Participants then attempted data manipulation and analysis tasks (early versions of the tasks used in the main study). The sessions concluded with a semi-structured interview discussing their experience with the prototype, what they liked, and what to improve upon. Sessions were conducted remotely through Microsoft Teams and lasted around 30 minutes each. They were recorded and transcribed. One of the authors analysed the transcripts with an informal open-coding approach [15], which included discussions with the research team. Our key findings are presented below.

### C.3 Results

Participants appreciated the ability to perform spreadsheet tasks with natural language queries rather than developing the correct formula, whether through personal experience or complex web searching and sensemaking, claiming that it was much more efficient (when it worked) and reduced the need for *"googling"* (FP2). However, they encountered several significant challenges in the process:

*C.3.1 It was hard for participants to recover from system errors.* Sometimes when the system output was obviously incorrect or the system did something unexpected, participants felt directionless about their options for fixing the errors. On one hand, there often existed a gap between the natural language intent (for example, *"add space flight and space walks"*) and the system output manifested in the spreadsheet (in this case, two new columns, one calculating "space flight = space flights * space flight (hr)", the other calculating "space walk = space walks * space walk (hr))", and it was difficult for participants to form hypotheses of what the root cause was without any knowledge of the logic generated by Codex. Even if participants had some intuitions on potential fixes, those intuitions were not systematic, and were often random guesses, such as *"should I be careful about pluralization or cases"* (FP5).

We observed that when an error occurred, participants only tried to reformulate their natural language intent twice on average before abandoning the system. When asked about what they would like to do next to solve the tasks, participants reported wanting to *"just figure out how to write a formula"* (FP4) or consult *"google"* (FP1).

*C.3.2 It was difficult for participants to develop mental models of the capabilities of the system.* Even after repeated use, participants still had poor notions about what the system could do, and how to phrase their queries to get the system to work reliably. Instead, their queries (and subsequent fixes, if applicable) were biased by their personal beliefs, speculations, and haphazard experimentation, such as the example discussed in section C.3.1. Participants wished to be able to better grasp the kinds of utterances, grammar, vocabulary, and level of specificity that is effective at generating the desired output. For example, FP4 wondered *"what does it [the system] understand? Do I have to use the same column names [as the ones in the table]?"*

---

[6]https://support.microsoft.com/en-us/office/analyze-data-in-excel-3223aab8-f543-4fda-85ed-76bb0295ffc4
[7]https://support.google.com/docs/answer/6280499

Moreover, participants found it difficult to generalise from a successful solution, to replicate the same or similar computation on different problems, since they could not identify *why* a query had been successful (e.g., was it their word choice, level of abstraction or problem decomposition, specific operations, or mentioning exact elements in the spreadsheet?). They wished that the system could *"talk to me in English and say what it's doing"* (FP5), and felt that there was a missed opportunity of *"learning"* to *"make it [the same type of data manipulation] faster elsewhere"* (FP2).

*C.3.3 Participants expressed a lack of trust towards system results.* As opposed to spreadsheet formulas, which many spreadsheet users are familiar with, participants expressed frustration and a sense of distrust towards the instability of the results produced by their natural language queries. For example, FP2 complained that *"non-determinism is a pain"* when comparing the system behaviors with regular spreadsheet formulas that he could understand and obtain anticipated results from. In addition, when system produced some partial instead of expected results, participants were not sure if they were *"on the right track"* (FP3). Last but not least, participants thought that, unlike formulas, the current natural language interface lacked a way for them to effectively verify the correctness of the system output. Though it was relatively easy to *"sanity check"* (FP4) a few *"instances or examples [of the system output]"* during the study, participants felt such spot checking approach was insufficient and intractable in reality, especially for large data tables with *"perhaps thousands or tens of thousands of rows"* (FP1).

**In summary**, our exploratory study gave evidence in our specific application context both of the abstraction matching problem as well as the issues around error recovery, mental models, and trust identified by Sarkar et al. [95] and Srinivasa Ragavan et al. [105]. This motivated the development of the grounded abstraction matching approach.

# D STUDY TASKS

## D.1 Task descriptions shown to participants

- Task 1: *As you could probably tell from the spreadsheet, each city has its own home team, for example the city of Denver has Broncos and Baltimore has Ravens. Your task is: Use the sidebar to find out the number of superbowls the city of New Orleans has won?*
- Task 2: *This is a worksheet containing the details of some of the NASA astronauts who have been into space, such as their name, birth place, hours of space flight, as well as the space missions (separated by ",") that they have flown before. Depending on the primary objectives, the duration of each space mission can vary a lot. Now, please imagine that you would like to find out on average how long each mission is for every astronaut. Your task is to create a new column that calculates that.*
- Task 3: *This is a worksheet listing the prices of some houses for sale as well as their details, such as the number of bedrooms, bathrooms, where it is located, etc. Now, imagine that you would like to check if a house if relatively new. In addition, you would like to see if it has a basement that you can use as storage, and if is has been renovated before. For the sake of this task, new houses are those built in or after 1970. Your task is to use the sidebar to create a column checking if a house satisfies all three of the aforementioned criteria.*

## D.2 Task data

Table 12: Full data table for task 1 (Super Bowl) in the study as described in section 5.

| Date | Winner | Winner Pts | Loser | Loser Pts | MVP | Stadium | Host City | Host State |
|---|---|---|---|---|---|---|---|---|
| Feb 2 2020 | Kansas City Chiefs | 31 | San Francisco 49ers | 20 | Patrick Mahomes | Hard Rock Stadium | Miami | Florida |
| Feb 3 2019 | New England Patriots | 13 | Los Angeles Rams | 3 | Julian Edelman | Mercedes-Benz Stadium | Atlanta | Georgia |
| Feb 4 2018 | Philadelphia Eagles | 41 | New England Patriots | 33 | Nick Foles | U.S. Bank Stadium | Minneapolis | Minnesota |
| Feb 5 2017 | New England Patriots | 34 | Atlanta Falcons | 28 | Tom Brady | NRG Stadium | Houston | Texas |
| Feb 7 2016 | Denver Broncos | 24 | Carolina Panthers | 10 | Von Miller | Levi's Stadium | Santa Clara | California |
| Feb 1 2015 | New England Patriots | 28 | Seattle Seahawks | 24 | Tom Brady | University of Phoenix Stadium | Glendale | Arizona |
| Feb 2 2014 | Seattle Seahawks | 43 | Denver Broncos | 8 | Malcolm Smith | MetLife Stadium | East Rutherford | New Jersey |
| Feb 3 2013 | Baltimore Ravens | 34 | San Francisco 49ers | 31 | Joe Flacco | Mercedes-Benz Superdome | New Orleans | Louisiana |
| Feb 5 2012 | New York Giants | 21 | New England Patriots | 17 | Eli Manning | Lucas Oil Stadium | Indianapolis | Indiana |
| Feb 6 2011 | Green Bay Packers | 31 | Pittsburgh Steelers | 25 | Aaron Rodgers | Cowboys Stadium | Arlington | Texas |
| Feb 7 2010 | New Orleans Saints | 31 | Indianapolis Colts | 17 | Drew Brees | Sun Life Stadium | Miami | Florida |
| Feb 1 2009 | Pittsburgh Steelers | 27 | Arizona Cardinals | 23 | Santonio Holmes | Raymond James Stadium | Tampa | Florida |
| Feb 3 2008 | New York Giants | 17 | New England Patriots | 14 | Eli Manning | University of Phoenix Stadium | Glendale | Arizona |
| Feb 4 2007 | Indianapolis Colts | 29 | Chicago Bears | 17 | Peyton Manning | Dolphin Stadium | Miami | Florida |
| Feb 5 2006 | Pittsburgh Steelers | 21 | Seattle Seahawks | 10 | Hines Ward | Ford Field | Detroit | Michigan |
| Feb 6 2005 | New England Patriots | 24 | Philadelphia Eagles | 21 | Deion Branch | Alltel Stadium | Jacksonville | Florida |
| Feb 1 2004 | New England Patriots | 32 | Carolina Panthers | 29 | Tom Brady | Reliant Stadium | Houston | Texas |
| Jan 26 2003 | Tampa Bay Buccaneers | 48 | Oakland Raiders | 21 | Dexter Jackson | Qualcomm Stadium | San Diego | California |
| Feb 3 2002 | New England Patriots | 20 | St. Louis Rams | 17 | Tom Brady | Louisiana Superdome | New Orleans | Louisiana |
| Jan 28 2001 | Baltimore Ravens | 34 | New York Giants | 7 | Ray Lewis | Raymond James Stadium | Tampa | Florida |
| Jan 30 2000 | St. Louis Rams | 23 | Tennessee Titans | 16 | Kurt Warner | Georgia Dome | Atlanta | Georgia |
| Jan 31 1999 | Denver Broncos | 34 | Atlanta Falcons | 19 | John Elway | Pro Player Stadium | Miami | Florida |
| Jan 25 1998 | Denver Broncos | 31 | Green Bay Packers | 24 | Terrell Davis | Qualcomm Stadium | San Diego | California |
| Jan 26 1997 | Green Bay Packers | 35 | New England Patriots | 21 | Desmond Howard | Louisiana Superdome | New Orleans | Louisiana |

Table 13: Full data table for task 2 (astronauts) in the study as described in section 5.

| Name | Status | Birth Date | Birth Place | Gender | Space Flight (hr) | Space Walks | Space Walks (hr) | Missions |
|---|---|---|---|---|---|---|---|---|
| Joseph M. Acaba | Active | 5/17/67 | Inglewood, CA | Male | 3307 | 2 | 13 | STS-119 (Discovery), ISS-31/32 (Soyuz) |
| Loren W. Acton | Retired | 7/3/36 | Lewiston, MT | Male | 190 | 0 | 0 | STS 51-F (Challenger) |
| James C. Adamson | Retired | 3/3/46 | Warsaw, NY | Male | 334 | 0 | 0 | STS-28 (Columbia), STS-43 (Atlantis) |
| Thomas D. Akers | Retired | 5/20/51 | St. Louis, MO | Male | 814 | 4 | 29 | STS-41 (Discovery), STS-49 (Endeavor), STS-61 (Endeavor), STS-79 (Atlantis) |
| Buzz Aldrin | Retired | 1/20/30 | Montclair, NJ | Male | 289 | 2 | 8 | Gemini 12, Apollo 11 |
| Andrew M. Allen | Retired | 4/8/55 | Philadelphia, PA | Male | 906 | 0 | 0 | STS-46 (Atlantis), STS-62 (Columbia), STS-75 (Columbia) |
| Joseph P. Allen | Retired | 6/27/37 | Crawsfordsville, IN | Male | 313 | 2 | 12 | ST-5 (Columbia), STS 51-A (Discovery) |
| Scott D. Altman | Retired | 8/15/59 | Lincoln, IL | Male | 1236 | 0 | 0 | STS-90 (Columbia), STS-106 (Atlantis), STS-109 (Columbia), STS-125 (Atlantis) |
| William A. Anders | Retired | 10/17/33 | Hong Kong | Male | 147 | 0 | 0 | Apollo 8 |
| Clayton C. Anderson | Retired | 2/23/59 | Omaha, NE | Male | 4005 | 6 | 38 | STS-117/120 (Atlantis/Discovery), STS-131 (Discovery) |
| Michael P. Anderson | Deceased | 12/25/59 | Plattsburgh, NY | Male | 594 | 0 | 0 | STS-89 (Endeavor), STS-107 (Columbia) |
| Dominic A. Antonelli | Active | 8/23/67 | Detroit, MI | Male | 579 | 0 | 0 | STS-119 (Discovery), STS-132 (Atlantis) |
| Jerome Apt III | Retired | 4/18/49 | Springfield, MA | Male | 847 | 2 | 11 | STS-37 (Atlantis), STS-47 (Endeavor), STS-59 (Endeavor), STS-79 (Atlantis) |
| Lee J. Archambault | Retired | 8/25/60 | Oak Park, IL | Male | 639 | 0 | 0 | STS-117 (Atlantis), STS-119 (Discovery) |
| Neil A. Armstrong | Deceased | 5/8/30 | Wapakoneta, OH | Male | 205 | 1 | 2 | Gemini 8, Apollo 11 |
| Richard R. Arnold II | Active | 11/26/63 | Cheverly, MD | Male | 307 | 2 | 12 | STS-119 (Discovery) |
| Jeffrey S. Ashby | Retired | 1/6/54 | Dallas, TX | Male | 655 | 0 | 0 | STS-93 (Columbia), STS-100 (Endeavor), STS-112 (Atlantis) |
| James P. Bagian | Retired | 2/22/52 | Philadelphia, PA | Male | 337 | 0 | 0 | STS-29 (Discovery), STS-40 (Columbia) |
| Ellen S. Baker | Retired | 4/27/53 | Fayetteville, NC | Female | 686 | 0 | 0 | STS-34 (Atlantis), STS-50 (Columbia), STS-71 (Atlantis) |
| Michael A. Baker | Management | 10/27/53 | Memphis, TN | Male | 965 | 0 | 0 | STS-43 (Atlantis), STS-52 (Columbia), STS-68 (Endeavor), STS-81 (Atlantis) |
| Michael R. Barratt | Active | 4/16/59 | Vancouver, WA | Male | 5075 | 1 | 5 | ISS-19/20 (Soyuz), STS-133 (Discovery) |
| Daniel T. Barry | Retired | 12/30/53 | Norwalk, CT | Male | 733 | 4 | 26 | STS-72 (Endeavor), STS-96 (Discovery), STS-105 (Discovery) |
| John-David F. Bartoe | Retired | 11/17/44 | Abington, PA | Male | 190 | 0 | 0 | STS 51-F (Challenger) |

Table 14: Full data table for task 3 (houses) in the study as described in section 5.

| price | bedrooms | bathrooms | floors | sqft_above | sqft_basement | yr_built | yr_renovated | zipcode | latitude | longitude |
|---|---|---|---|---|---|---|---|---|---|---|
| 221900 | 3 | 1 | 1 | 1180 | 0 | 1955 | 0 | 98178 | 47.5112 | -122.257 |
| 538000 | 3 | 2.25 | 2 | 2170 | 400 | 1951 | 1991 | 98125 | 47.721 | -122.319 |
| 180000 | 2 | 1 | 1 | 770 | 0 | 1933 | 0 | 98028 | 47.7379 | -122.233 |
| 604000 | 4 | 3 | 1 | 1050 | 910 | 1965 | 1998 | 98136 | 47.5208 | -122.393 |
| 510000 | 3 | 2 | 1 | 1680 | 0 | 1987 | 0 | 98074 | 47.6168 | -122.045 |
| 1230000 | 4 | 4.5 | 1 | 3890 | 1530 | 2001 | 0 | 98053 | 47.6561 | -122.005 |
| 257500 | 3 | 2.25 | 2 | 1715 | 0 | 1995 | 0 | 98003 | 47.3097 | -122.327 |
| 291850 | 3 | 1.5 | 1 | 1060 | 0 | 1963 | 0 | 98198 | 47.4095 | -122.315 |
| 229500 | 3 | 1 | 1 | 1050 | 730 | 1960 | 2010 | 98146 | 47.5123 | -122.337 |
| 323000 | 3 | 2.5 | 2 | 1890 | 0 | 2003 | 0 | 98038 | 47.3684 | -122.031 |
| 662500 | 3 | 2.5 | 1 | 1860 | 1700 | 1965 | 1998 | 98007 | 47.6007 | -122.145 |
| 468000 | 2 | 1 | 1 | 860 | 300 | 1942 | 0 | 98115 | 47.69 | -122.292 |
| 310000 | 3 | 1 | 1.5 | 1430 | 0 | 1927 | 0 | 98028 | 47.7558 | -122.229 |
| 400000 | 3 | 1.75 | 1 | 1370 | 400 | 1977 | 2001 | 98074 | 47.6127 | -122.045 |
| 530000 | 5 | 2 | 1.5 | 1810 | 0 | 1900 | 0 | 98107 | 47.67 | -122.394 |
| 650000 | 4 | 3 | 2 | 1980 | 970 | 1979 | 0 | 98126 | 47.5714 | -122.375 |
| 395000 | 3 | 2 | 2 | 1890 | 0 | 1994 | 0 | 98019 | 47.7277 | -121.962 |
| 485000 | 4 | 1 | 1.5 | 1600 | 0 | 1916 | 0 | 98103 | 47.6648 | -122.343 |
| 189000 | 2 | 1 | 1 | 1200 | 0 | 1921 | 0 | 98002 | 47.3089 | -122.21 |
| 230000 | 3 | 1 | 1 | 1250 | 296 | 1970 | 2019 | 98003 | 47.3343 | -122.306 |
| 385000 | 4 | 1.75 | 1 | 860 | 760 | 1947 | 0 | 98133 | 47.7025 | -122.341 |
| 2000000 | 3 | 2.75 | 1 | 2330 | 720 | 1968 | 0 | 98040 | 47.5316 | -122.233 |
| 285000 | 5 | 2.5 | 2 | 2270 | 200 | 1995 | 2015 | 98092 | 47.3266 | -122.169 |
| 252700 | 2 | 1.5 | 1 | 1070 | 0 | 1985 | 0 | 98030 | 47.3533 | -122.166 |

# E EXTENDED QUERY EPISODE EXAMPLES
## E.1 Extended examples for failure modes

### Table 15: Examples of input failures encountered by participants

| Failure Mode | Example Query (Participant-Task) | Generated Python Code | Explanation of Failure |
| --- | --- | --- | --- |
| **Wrong input** | *How many times has the city of New Orleans appear in the winners column* (P24-1) | `df[df['Host City'] == 'New Orleans'].shape[0]` | The model should have used the "Winner" column instead of the "Host City" column. |
| **Soft wrong input** | *how many superbowls has the city of New Orleans won* (P27-1) | `df[df['Host City'] == 'New Orleans'].shape[0]` | The model should have used the "Winner" column instead of the "Host City" column. |

### Table 16: Examples of output failures encountered by participants

| Failure Mode | Example Query (Participant-Task) | Generated Python Code | Explanation of Failure |
| --- | --- | --- | --- |
| **Overwrite** | *Create col showing how many missions* (P1-2) | `df['Missions'] = df['Missions'].str.split(',')`<br>`df['Missions'] = df['Missions'].str.len()` | The model should not have overwritten the "Missions" column in both lines. |
| **Soft overwrite** | *Split the Missions column and count the number of splits* (P28-2) | `df['Missions'] = df['Missions'].str.split(', ')`<br>`df['Mission Count'] = df['Missions'].str.len()` | The model should not have overwritten the "Missions" column in the first line of code. |
| **Extra columns** | *Add a column to see if the house has basement, been renovated and was built later than 1970* (P5-3) | `df['has_basement'] = df['sqft_basement'] > 0`<br>`df['has_renovated'] = df['yr_renovated'] > 0`<br>`df['built_after_1970'] = df['yr_built'] > 1970` | The model should not have generated the extra columns of "has_basement", "has_renovated", or "built_after_1970", since the participant specifically asked for only one column. |
| **Missing columns** | *(1) Add new column with count number of missions, each separated by a ',', (2) Add new column and calculate average space flight (hr) per name per mission. Return type: new columns of number* (P7-2) | `df['Mission Count'] = df['Missions'].str.split(',').str.len()` | The model failed to create a second column that the participant asked for that should "calculate average space flight (hr) per name per mission." |

### Table 17: Examples of logic failures encountered by participants

| Failure Mode | Example Query (Participant-Task) | Generated Python Code | Explanation of Failure |
| --- | --- | --- | --- |
| **Partial answer** | *Returning the time of "New Orleans" appears in the winner column* (P16-1) | `df[df['Winner'].str.contains('New Orleans')]` | The model only created a new table that selected the rows where "New Orleans" appears in the "Winner" column. However, the user can easily achieve the final goal by adding a new step to "return the number of rows" from the new table. |
| **Raw data output** | *New column number of missions* (P20-2) | `df['Number of Missions'] = [2, 1, 2, 4, 2, 3, 2,`<br>`↪ 4, 1, 3, 2, 2, 2, 4, 2, 1, 3, 2, 3, 4, 2, 2,`<br>`↪ 1]` | The model hallucinated data values in the "Number of Missions" column rather than generating code to calculate those values. |
| **Wrong heuristic** | *List all the teams that have lost to New England Patriots* (P21-1) | `df[df['Loser'] == 'New England Patriots']` | The model should have used the "Winner" column instead of the "Loser" column to perform the filtering. |
| **Other incorrect** | *Return 3 columns for if year built >=1970, basement >0, renovated >0* (P11-3) | `def get_features(df):`<br>`  df['built_after_1970'] = df['yr_built'] >= 1970`<br>`  df['has_basement'] = df['sqft_basement'] > 0`<br>`  df['has_renovated'] = df['yr_renovated'] > 0`<br>`  return df[['built_after_1970', 'has_basement',`<br>`↪ 'has_renovated']]` | The model generated some code defining a `get_features` function, but failed to include any code that runs it, nor was the logic of the function syntactically correct. |

## E.2 Extended examples for rewrite strategies

Table 18: Examples of scoping changes. Orange and green highlights indicate differences (removals and additions, respectively) between a query and its follow-up query (the same color scheme applies to Table 19-21 as well).

| Rewrite Strategy | Previous Query (Participant-Task) **Generated Python Code** **Grounded Utterance** | Follow-up Query **Generated Python Code** **Grounded Utterance** |
|---|---|---|
| **Add steps** | *Define mission_count by splitting Missions by ','* (P6-2) `df['mission_count'] = df['Missions'].str.split(',').str.len()` (1) create column mission_count, (2) select column Missions, (3) split the text on ',', (4) len | *Define mission_count by splitting Missions by ',', then divide Space Flight by mission_count* `df['mission_count'] = df['Missions'].str.split(',').str.len()` `df['Space Flight (hr)'] = df['Space Flight (hr)'] / df['mission_count']` (1) create column mission_count from len from the text split on ',' from column Missions, (2) create column Space Flight (hr) from column Space Flight (hr) divided by column mission_count. |
| **Reduce scope** | *Create column good where year built is greater than or equal to 1970 AND squarefoot basement is not 0 and year renovated is not 0* (P9-3) `df['good'] = ((df['yr_built'] >= 1970) & (df['sqft_basement'] != 0) & (df['yr_renovated'] != 0))` (1) create column good, (2) column yr_built greater than or equal to 1970 and column sqft_basement NotEq 0 and column yr_renovated NotEq 0. | *Create column good where year built is greater than or equal to 1970* `df['good'] = df['yr_built'] >= 1970` (1) create column good, (2) column yr_built greater than or equal to 1970. |

Table 19: Examples of elaboration changes.

| Rewrite Strategy | Previous Query (Participant-Task)<br>Generated Python Code<br>Grounded Utterance | | Follow-up Query<br>Generated Python Code<br>Grounded Utterance |
|---|---|---|---|
| Elaborate how | *Add a column of the average flight hour of each mission for each astronaut* (P5-2)<br><br>```python
df['Average Flight Hour'] = [1653.5, 190, 167, 407, 289, 302,
    313, 309, 147, 1001.25, 297, 289.5, 423.5, 319.5, 205,
    307, 327.5, 168.5, 343, 482.5, 2537.5, 366.5, 190]
```<br><br>N/A | ⇒ | *Add a column of the value of total space flight hour divided by the number of missions for each astronaut*<br><br>```python
df['Average Mission Time'] = df['Space Flight (hr)'] /
    df['Missions'].str.count('\(')
```<br><br>(1) create column Average Mission Time,<br>(2) column Space Flight (hr) divided by count \( from column Missions. |
| Elaborate what | *how many superbowls has the city of New Orleans won* (P27-1)<br><br>```python
df[df['Host City'] == 'New Orleans'].shape[0]
```<br><br>(1) select rows where column Host City is New Orleans,<br>(2) return number of rows. | ⇒ | *how many superbowls has New Orleans Saints won*<br><br>```python
df[df['Winner'] == 'New Orleans Saints'].shape[0]
```<br><br>(1) select rows where column Winner is New Orleans Saints,<br>(2) return number of rows. |
| Name output columns | *Return true if year built >=1970 AND basement >0 and renovated TRUE* (P11-3)<br><br>```python
def is_renovated(row):
    if row['yr_built'] >= 1970 and row['sqft_basement'] > 0
        and row['yr_renovated'] > 0:
        return True
    else:
        return False
df['renovated'] = df.apply(is_renovated, axis=1)
print(df)
```<br><br>N/A | ⇒ | *Return review column where if year built >=1970 AND basement >0 AND renovated >0*<br><br>```python
def review(row):
    if row['yr_built'] >= 1970 and row['sqft_basement']
        > 0 and row['yr_renovated'] > 0:
        return 1
    else:
        return 0
df['review'] = df.apply(review, axis=1)
print(df)
```<br><br>N/A |
| Specify input columns | *How many super bowls has New Orleans won* (P9-1)<br><br>```python
df[df['Host City'] == 'New Orleans']['Winner'].count()
```<br><br>(1) select rows where column Host City is New Orleans,<br>(2) select column Winner,<br>(3) count. | ⇒ | *Select column "winner" where text includes new orleans*<br><br>```python
df[df['Winner'].str.contains('New Orleans')]
```<br><br>(1) select rows where contains 'New Orleans' from column Winner. |
| Specify output type | *Select rows where column yr_built greater than 1970 and column yr_renovated NptEq 0 and column sqft_basement NotEq 0* (P16-3)<br><br>```python
df[(df['yr_built'] > 1970) & (df['yr_renovated'] != 0) &
    (df['sqft_basement'] != 0)]
```<br><br>(1) select rows where column yr_built greater than 1970 and column yr_renovated NotEq 0 and column sqft_basement NotEq 0. | ⇒ | *Create a column where column yr_built greater than 1970 and column yr_renovated NotEq 0 and column sqft_basement NotEq 0*<br><br>```python
df['new_house'] = df.apply(lambda row: 1 if
    row['yr_built'] > 1970 and row['yr_renovated'] != 0
    and row['sqft_basement'] != 0 else 0, axis=1)
```<br><br>N/A |



Table 20: Examples of language restructuring changes.

| Rewrite Strategy | Previous Query (Participant-Task) **Generated Python Code** **Grounded Utterance** | | Follow-up Query **Generated Python Code** **Grounded Utterance** |
|---|---|---|---|
| **Self breakdown** | *Count the number of rows with 'New Orleans' in the winner column* (P12-1) `df[df['Winner'] == 'New Orleans'].count()` N/A | ⇒ | *(1) Create a new column called city that drops the last word in the winner column, (2) Count the number of rows in city that say "New Orleans"* `df['city'] = df['Winner'].str.split().str[-1]` `df['city'].value_counts()` N/A |
| **Reuse of the system breakdown** | *The number of superbowls the city of New Orleans has won* (P5-1) `df[df['Host City'] == 'New Orleans']['Winner'].count()` (1) select rows where column Host City is New Orleans, (2) select column Winner, (3) count. | ⇒ | *(1) select rows where column Winner is New Orleans Saints, (2) count* `df[df['Winner'] == 'New Orleans Saints'].count()` (1) select rows where column Winner is New Orleans Saints, (2) count. |
| **System-like** | *how many super bowls has New Orleans won* (P9-1) `df[df['Host City'] == 'New Orleans']['Winner'].count()` (1) select rows where column Host City is New Orleans (2) select column Winner (3) count | ⇒ | *select column winner where text includes new orleans* `df[df['Winner'].str.contains('New Orleans')]` (1) select rows where contains New Orleans from column Winner |
| **Code-like syntax** | *Create a column that shows the value of Space Flight (hr) divided by the number of items of the Missions column* (P26-2) `df['Space Flight (hr) per Mission'] = df['Space Flight`↪ `(hr)'] / df['Missions'].str.count(',') + 1` (1) create column Space Flight (hr) per Mission, (2) column Space Flight (hr) divided by count "," from column Missions + 1. | ⇒ | *(1) create column Space Flight (hr) per Mission, (2) column Space Flight (hr) divided by (count , from column Missions + 1)* `df['Space Flight (hr) per Mission'] = df['Space Flight (hr)']`↪ `/ (df['Missions'].str.count(',') + 1)` (1) create column Space Flight (hr) per Mission, (2) column Space Flight (hr) divided by count "," from column Missions + 1. |



Table 21: Examples of intent shaping changes.

| Rewrite Strategy | Previous Query (Participant-Task)<br>Generated Python Code<br>Grounded Utterances | | Follow-up Query<br>Generated Python Code<br>Grounded Utterances |
|---|---|---|---|
| New intent | *Select rows where basement > 0 and yr_built >= 1970 and yr_renovated > 0* (P21-3)<br><br>`df[(df['sqft_basement'] > 0) & (df['yr_built'] >= 1970) &`<br>`↪ (df['yr_renovated'] > 0)]`<br><br>N/A | ⇒ | *Create a column called consider? where the value is true if basement > 0 and yr_built >= 1970 and yr_renovated > 0*<br><br>`"df['consider?'] = df['sqft_basement'] > 0`<br>`df['consider?'] = df['consider?'] & (df['yr_built'] >= 1970)`<br>`df['consider?'] = df['consider?'] & (df['yr_renovated'] > 0)"`<br><br>N/A |
| Start over | *Use winner column to subtract the winner team column to get the winner city* (22-1)<br><br>`df['Winner City'] = df['Winner'] - df['Winner team']`<br><br>(1) create column Winner City,<br>(2) column Winner - column Winner team. | ⇒ | *Create a new winner city that remove the winner team from the winner column*<br><br>`df['Winner City'] = df['Winner'].str.replace(r'\b\w+\b', '')`<br><br>(1) create column Winner City,<br>(2) select column Winner,<br>(3) replace '\b\w+\b' with ''. |
| Next step | *Count the number of mission in column I delimited by comma* (P24-2)<br><br>`df['Mission Count'] = df['Missions'].str.count(',') + 1`<br><br>(1) create column Mission Count,<br>(2) count ',' from column Missions + 1. | ⇒ | *For each name calculate the hours of space flight divided by the mission count*<br><br>`df['Hours per Mission'] = df['Space Flight (hr)'] /`<br>`↪ df['Mission Count']`<br><br>(1) create column Hours per Mission,<br>(2) column Space Flight (hr) divided by column Mission Count. |
| Testing | *Create a new column that counts how many strings are separated by ',' in the "Missions" column. Create another column that calculates "Space Flight (hr)' divided by "Missions_Count"* (P25-2)<br><br>`df['Missions_Count'] = df['Missions'].str.count(',') + 1`<br>`df['Space Flight (hr) per Mission'] = df['Space Flight`<br>`↪ (hr)'] / df['Missions_Count']`<br><br>N/A | ⇒ | *new column: # of strings separated by ',' in "Missions". Create another column that calculates "Space Flight (hr)' divided by "Missions_Count"*<br><br>`df['Missions_Count'] = df['Missions'].str.count(',') + 1`<br>`df['Avg_Mission_Length'] = df['Space Flight (hr)'] /`<br>`↪ df['Missions_Count']`<br><br>N/A |